\documentclass{article}

\usepackage{pslatex}
\usepackage{epsfig}
\usepackage{newlfont}

\begin{document}

\thispagestyle{empty}

\phantom{x}
\vspace{-2cm}
\hfill TTP00-12

\hfill NIKHEF-2000-017

\hfill hep-ph/0007221

\hfill July 2000

\begin{center}
{\Large\bf Parallelizing the Symbolic Manipulation Program FORM\\
Part I: Workstation Clusters \& Message Passing}
\end{center}
\begin{center}
{\Large D.Fliegner$^{\dagger}$, 
A.R\'etey$^{\dagger}$, J.A.M. Vermaseren$^{\ddagger}$}
\end{center}
\noindent
$^{\dagger}$Institut f\"ur Theoretische Teilchenphysik,
Universit\"at Karlsruhe, D-76128 Karlsruhe, Germany\\
$^{\ddagger}$NIKHEF, P.O. Box 41882, 1009 DB, Amsterdam, The Netherlands
\vspace{3mm}

\begin{center}
{\large\bf Abstract}
\end{center}

\noindent
The present paper is the first of a series of papers reporting on the 
parallelization of the symbolic manipulation program FORM on different
parallel architectures. Part I deals with workstation clusters using
dedicated network hardware and the messages passing libraries (MPI and
PVM). After a short introduction to the sequential version of FORM a
detailed analysis of the different platforms used is given and the
structure of the parallel version of FORM is explained. The forthcoming
part II will describe the parallelization of FORM on SMP (symmetrical 
multi-processing) architectures.

\begin{center}
keywords: symbolic manipulation, computer algebra, high energy physics,
perturbation theory, FORM
\end{center}

\vspace{5mm}
\noindent
{\large\bf 1. The Sequential Version of FORM}
\vspace{5mm}

\noindent
{\large\bf 1.1 Introduction}
\vspace{5mm}


\noindent
FORM is a program for symbolic manipulation of algebraic expressions 
specialized to handle very large expressions of millions of terms in an 
efficient and reliable way. Although it supplies a programming 
language that allows for the formulation of a wide area of symbolic 
algorithms it is currently mainly used for computations in high energy 
physics.
Especially in the field of perturbative calculations of higher order 
corrections to quantum field theoretical quantities (the so-called
multi loop calculations) FORM has become a standard tool.

A systematic introduction to the FORM programming language is of course
far beyond the scope of this paper and the reader is referred to
\cite{form}. Our main concern here is the discussion of the internal
mechanisms of FORM that become important in its parallelization.

Since FORM was designed for solving large problems, it is used in an 
non-interactive way by supplying a program that is executed. The results
and typically also some statistics about the progress of the calculation
are printed to screen or redirected to a file. 
For the following it is instructive to consider an example of a very simple 
FORM program:

\begin{verbatim}
symbols x,a,b;

local expression = a*x + x^2;

id x = a + b;

.sort

if(count(b,1)==1);
  multiply 4*a/b;
endif;

print;
.end
\end{verbatim}

\newpage
\noindent
Unlike most of the other computer algebra systems FORM is type-oriented, 
so that a program starts with the declaration of variables that are to be
used, in this case the {\tt symbols} {\tt x}, {\tt a} and {\tt b}. Then 
the expression(s) that are dealt with have to be defined. In this case
{\tt expression} is defined as a {\tt local} expression. Expressions
are sums (sequences) of single terms and are (almost) always completely
expanded to yield a unique representation. The rest of a program contains
the {\it statements} that define what is to be done with the expressions,
in this case: the {\tt id}entification of all occurrences of {\tt x} in 
{\tt expression} by the sum of $a$ and $b$, multiplying all terms with
a linear factor of $b$ by $4a/b$ and printing the {\tt expression}. 
In general a  program is divided into so called modules that are terminated
with ``dot''-instructions that cause the execution of the module. Our 
simple example consists of only two modules. Consequently there are two
``dot''-instruction: a {\tt .sort} and a {\tt .end}. In both cases the
result is sorted. {\tt .end} additionally terminates the program.
Running the program produces the following output:

\begin{verbatim}
FORM version 3.-1(Mar  7 2000). Run at: Tue Mar  7 21:35:15 2000
    symbols x,a,b;
    
    local expression = a*x + x^2;
    
    id x = a + b;
    
    .sort

Time =       0.00 sec    Generated terms =          5
       expression        Terms in output =          3
                         Bytes used      =         52
    
    if(count(b,1)==1);
      multiply 4*a/b;
    endif;
    
    print;
    .end

Time =       0.00 sec    Generated terms =          3
       expression        Terms in output =          2
                         Bytes used      =         32

   expression =
      14*a^2 + b^2;
\end{verbatim}

\noindent
The special properties of FORM result from the fact that it basically 
allows {\it local} operations on {\it single} terms only. Examples of
local operations are: replacing parts of a term by another term or a
whole expression, multiplying a term by a certain expression. 
{\it Non-local} operations like replacing a sum of two terms in an 
expression by another term involve {\it more than one} term at a time
and are strictly forbidden. Only in
the sorting procedure at the end of the modules non-local operations
are performed, namely identifying equivalent terms and adding up their
prefactors. Together with a flexible pattern matcher this seemingly
strong limitation still allows the formulation of general and very 
efficient algorithms.  On the other hand it enables handling expressions
as ``streams'' of terms, that can be read sequentially (from memory or
a file) and be worked on one at a time. This forms the basis for a 
sophisticated memory management that allows to deal with expressions that
are bigger than the memory (RAM) available.  The restriction to local
operations obviously also allows parallelism.


\vspace{5mm}
\noindent
{\large\bf 1.2 Internal Data Representation}
\vspace{5mm}


\noindent
FORM uses the following building blocks for the internal representation of
expressions:

\begin{itemize}

\item variables: are the basic building blocks. They must have a type
(symbols, vectors, indices, functions, \ldots) and can have additional
properties (commutativity, symmetry w.r.t arguments, \ldots).

\item coefficients: can be arbitrary rational numbers. The corresponding
arbitrary long integer arithmetics is implemented in FORM.

\item terms: are products of variables with a coefficient. Functions can
have several arguments that in turn can be (sub-)expressions.  

\item expressions: are sums (=sequences) of terms.

\end{itemize}

\noindent
The internal representation of these building blocks are sequences of
integer numbers (words). Typically the word size is half the maximum
size for integers on a given architecture, allowing an efficient 
implementation of the integer arithmetics. A variable is represented 
by a sub-term, a sequence of words, whose first entry is a word 
describing the type, followed by a word giving the length of the
sequence, a word describing the position of the variable in the internal
symbol table and eventually the exponent. In the case of vectors and
functions this is followed by the corresponding argument fields.
The representation of the coefficients is particularly interesting:
they are read in reversed order, the last word giving the length $n$
of the coefficient (including this entry) and the sign of the coefficient.
Numerator and denominator by definition have the same length so that
the last word of a coefficient is always an odd number. The $(n-1)/2$
words in front represent the denominator, the $(n-1)/2$ words in front
of the denominator represent the numerator. 
 
\begin{figure}[h!]
\begin{center}
\epsfig{file=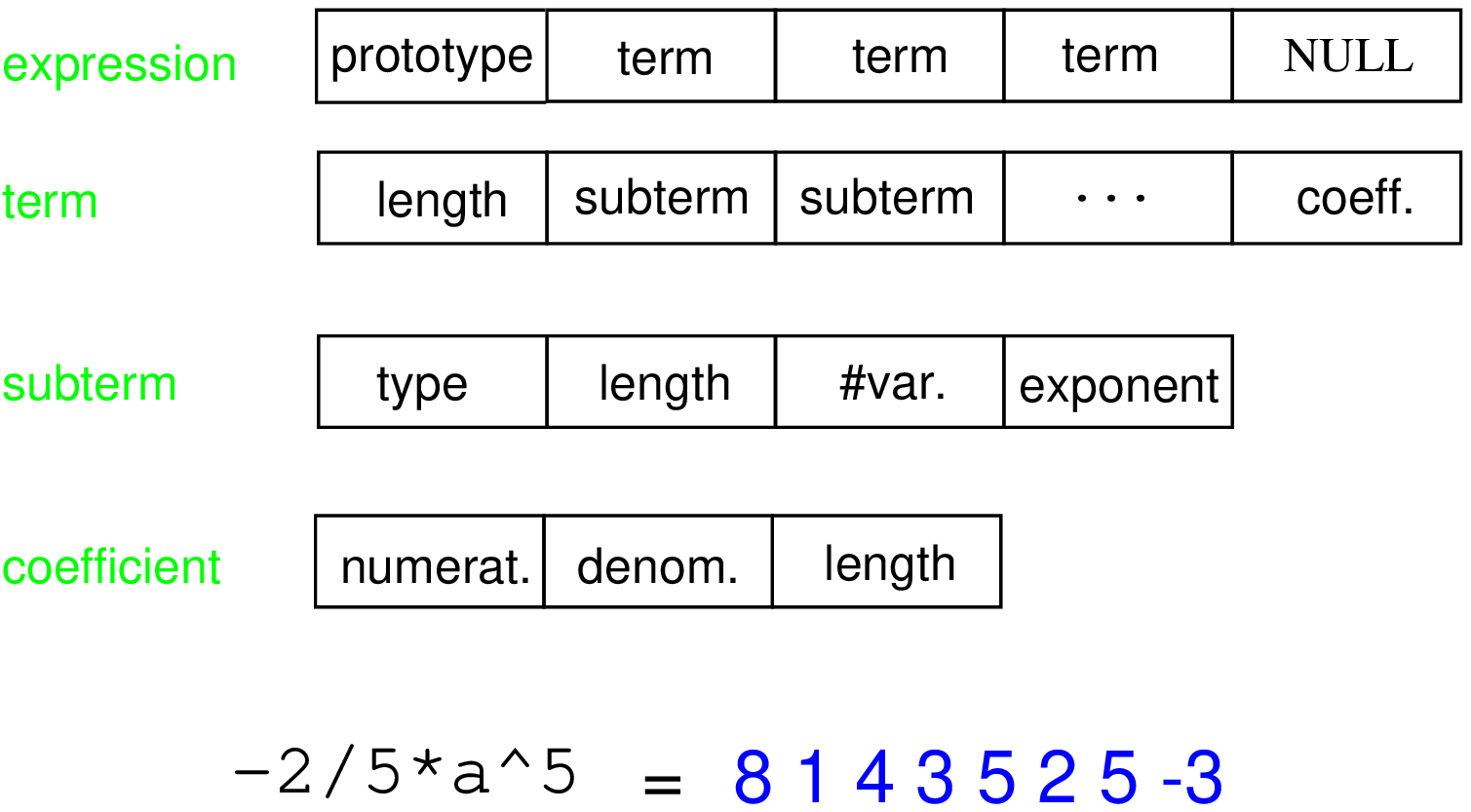,height=5cm}
\end{center}
\caption{Example for the internal representation of $-\frac{2}{5}a^5$
as a series of words (short integers). The first number gives the total
length (number of words) 8 of the term. The following numbers represent a 
subterm of type 1 (symbol) with total length 4. The number of the symbol
$a$ in the symbol table is 3 and the exponent is 5. The subterm is followed
by the coefficient with length 3 (absolute value of last number) and 
negative sign (sign of last number). The numerator is 2 and the denominator 5.
}
\label{intrep}
\end{figure}

\noindent
Figure \ref{intrep} shows a simple example for the internal representation. 
Using the (non-documented) statement {\tt write code;} it is possible to
have a look at the contents of the compiler buffers in the internal
representation. These buffers also contain constructs representing the
instructions and additional information needed for pattern matching that we
do not discuss here. Only the fact that the terms of an expression are
represented by sequences of words of known length is important for the
following.


\vspace{5mm}
\noindent
{\large\bf 1.3 Preprocessor and Compiler}
\vspace{5mm}


\noindent
As already mentioned FORM runs a program module by module. The execution
of the program is coordinated by the {\it preprocessor}. It merges the
program text from different sources (files), inserts the preprocessor
variables' values and evaluates the preprocessor instructions 
(lines beginning with a {\tt \#}). Moreover it passes the statements to the
compiler and eventually calls the processor in order to actually execute 
the modules when it encounters a dot-instruction.

\begin{figure}[h!]
\begin{center}
\epsfig{file=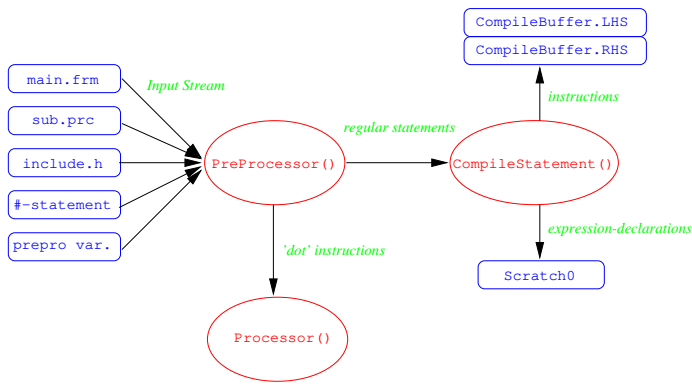,height=5cm}
\end{center}
\caption{The data flow during preprocessing and compiling.}
\label{flowchart1}
\end{figure}

The {\it compiler} translates the ASCII-input text line by line into the 
internal representation described in section 1.2. It writes the definitions 
of the expressions (the prototypes) into so called scratch-files (with 
extension {\tt .sc0} or {\tt .sc1}). The instructions are divided into left
hand sides (LHS) and right hand sides (RHS) and stored in the corresponding
buffers. For an {\tt id}-statement the LHS corresponds to the (sub-)term
that is to be replaced, the RHS corresponds to the expression it is to
be replaced with. For other statements the LHS contains other information
and in some cases the RHS might not even exist and there is a LHS only. 
The internal representation also contains all information necessary for
pattern matching. The whole procedure is pictured in figure \ref{flowchart1}.

\noindent
If the preprocessor encounters a dot-instruction it causes the execution
of the instructions stored in the compiler buffers by calling the function
{\tt Execute}, which in turn calls the function {\tt Processor}. For a given
expression the processor decides if it is going to be manipulated in the 
current module. If this is the case the corresponding terms are read from
the scratch-file by the function {\tt GetTerm} one at a time and passed to
the function {\tt Generator}. This function performs the actual execution of
the module for every single input term by generating the output terms
according to the instructions stored in the compiler buffers.


\vspace{5mm}
\noindent
{\large\bf 1.4 Generating Terms}
\vspace{5mm}


\noindent
The {\it generator} is the core of FORM. The complex recursive function 
{\tt Generator} takes a single term and performes the following operations:

\begin{itemize}

\item the term is being checked for sub-terms that have to be inserted.

\item the term is transformed to normal form by ordering the sub-terms.

\item the instructions of the compiler statement corresponding to the
given recursion depth are applied.

\item the result is brought to standard form by expanding the sub-terms.

\end{itemize}

\begin{figure}[h!]
\begin{center}
\epsfig{file=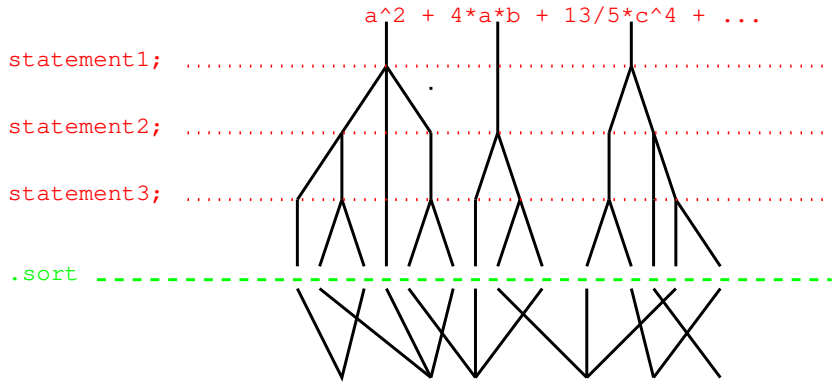,height=5cm}
\end{center}
\caption{Schematic picture of a generation tree.}
\label{generate}
\end{figure}

\noindent
Since the function {\tt Generator} is called recursively for a given term,
all the instructions in the compiler buffers are applied to a single term
before the generator proceeds with the next input term. A buffer called
{\tt workspace} is used as a stack for the recursion. The size of the 
work-space can be increased if necessary by providing a corresponding entry
in the set-file. Once all statements have been applied to the first term,
the next term in the workspace is taken. The generating procedure yields a 
tree-like structure for term generation (shown in figure \ref{generate}), 
where the execution is not line-oriented (as one might guess naively), but 
term-oriented.

\newpage
\noindent
{\large\bf 1.5 Sorting Terms}
\vspace{5mm}


\noindent
As described in the last section the generator produces a stream of output
terms for each input term given. Generally the output terms come unsorted
and in redundant form. In order to get a result in standard form, the output
terms have to be sorted and equivalent terms (terms equal up to a prefactor)
have to be summed up. This is done in a staged procedure using two buffers 
in the memory (the small buffer {\tt sBuffer} and the large buffer {\tt 
lBuffer}), whose sizes can be adjusted in the set-file and a temporary 
sort-file on disk. The unsorted terms are written into the small buffer
first. If the small buffer gets full, its content gets sorted and is
copied to the large buffer, freeing the small buffer for the next patch
of output terms. If the large buffer gets full, its content (pre-sorted
patches of terms) gets again sorted and is copied to the temporary file,
freeing the large buffer. If all terms have been generated for all 
expressions in a module, the terms that are still residing in the different
memory buffers and the sort-file have to be merged together to yield the
final expression. This is again been done stage by stage, if necessary
by using additional temporary sort-files. The resulting sorted stream of
terms is written to the output scratch-file, which is in turn used as
the input scratch-file for the next module.

\begin{figure}[h!]
\begin{center}
\epsfig{file=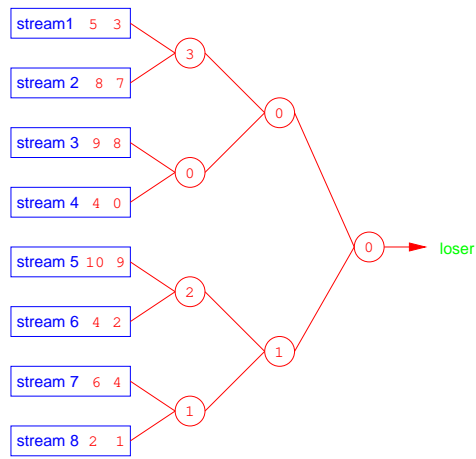,height=6cm}
\end{center}
\caption{The tree of losers algorithm for integers. A number of pre-sorted
integer streams are merged together in a tree-like structure, yielding the
so-called loser (the smallest element) at the root of the tree.}
\label{treeoflosers}
\end{figure}

\begin{figure}[h!]
\begin{center}
\epsfig{file=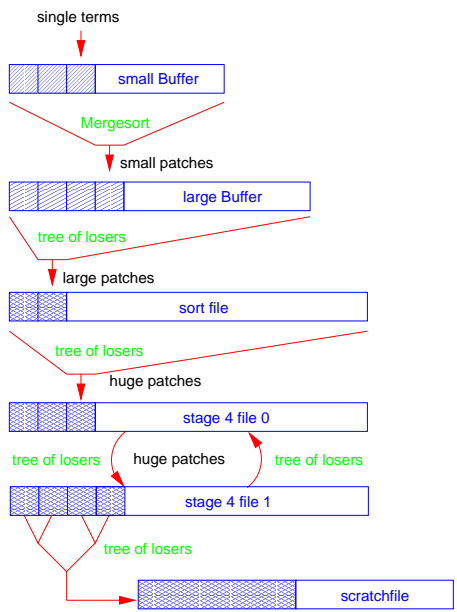,height=8cm}
\end{center}
\caption{The different stages of the sorting procedure.}
\label{stagesort}
\end{figure}

\noindent
For the sorting of the (unsorted) terms in the first small buffer a 
modified version of the mergesort-algorithm \cite{knuth} is used. Like for
the well known quicksort-algorithm the number of operations necessary for
sorting a problem of size (numbers of terms) $n$ grows like $n \log n$.
Quicksort is slightly faster, but mergesort provides higher stability
against worst cases and the implementation of cancellation of terms can 
be done more easily. Pre-sorted patches of terms are always merged together
by the ``tree of losers'' algorithm (the idea of the algorithm is shown
for integers in figure \ref{treeoflosers}), that is described in 
\cite{knuth} for file-to-file sorting. Figure \ref{stagesort} gives an 
overall view on the staged sorting procedure.

\begin{figure}[h!]
\begin{center}
\epsfig{file=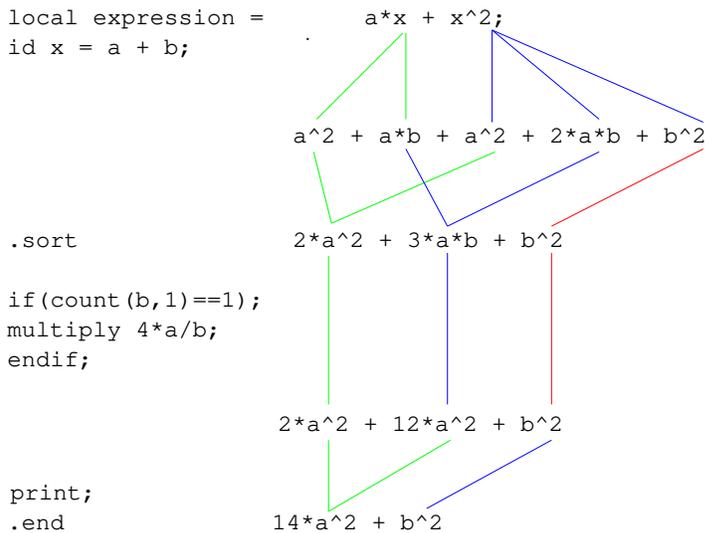,height=7cm}
\end{center}
\caption{Generating and sorting trees for the introductory example.}
\label{example}
\end{figure}

\noindent
Like the generation of terms the sorting of the output terms yields a tree
like structure. Figure \ref{example} shows the generating and sorting trees
for the example in the introduction.

\begin{figure}[h!]
\begin{center}
\epsfig{file=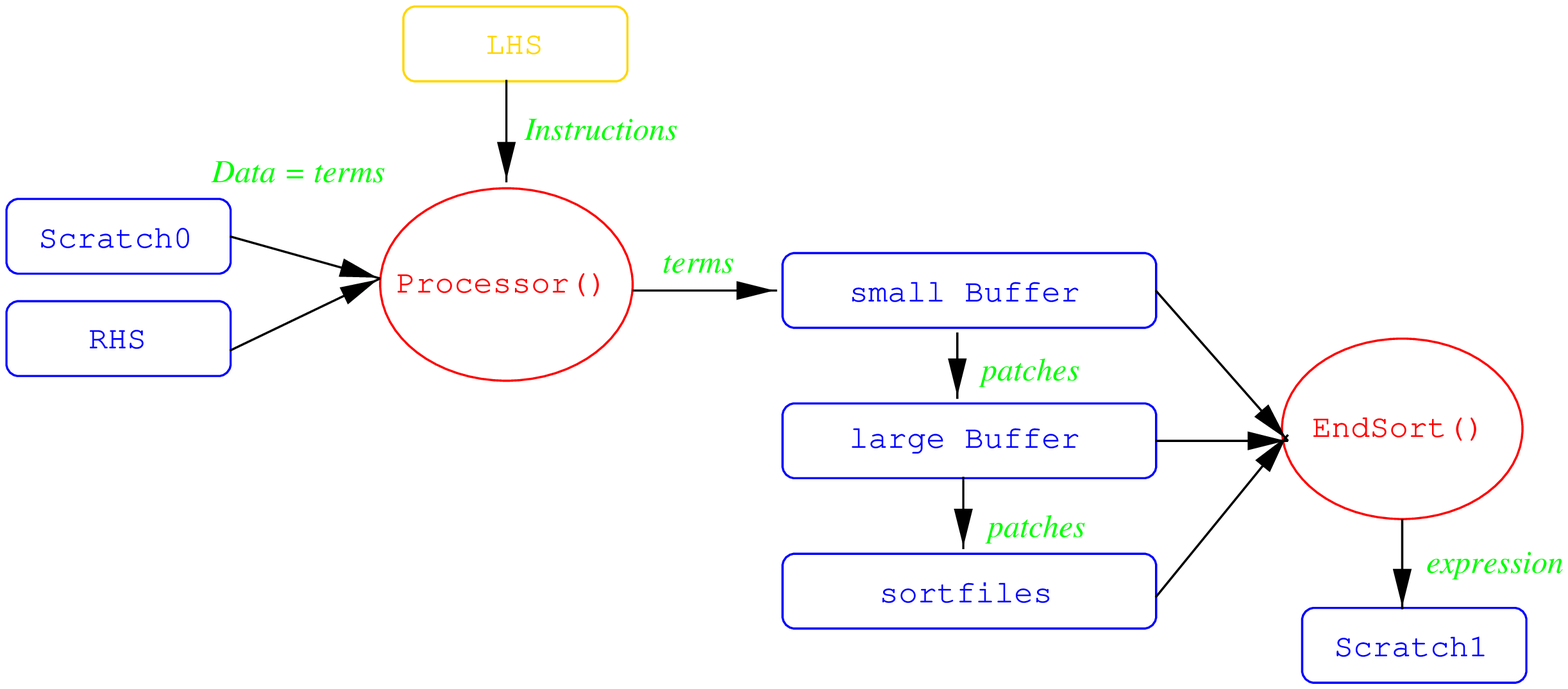,height=5cm}
\end{center}
\caption{The data flow during the generating and sorting phase.}
\label{flowchart2}
\end{figure}

The data flow for the generating and sorting phase of the execution of a
module is shown in \ref{flowchart2}.

\vspace{5mm}
\noindent
{\large\bf 1.6 Optimizing and Analyzing the Sequential Code}
\vspace{5mm}

Before any effort was made to parallelize FORM, the first step was 
to analyze and optimize the sequential code. FORM was ported to DEC
UNIX and the internal used word length was changed from 16 bits
(which results in 32-bit arithmetic operations) to 32 bits
(64-bit arithmetics), hereby using the full register width of the
Alpha-processors.
Afterwards this new 64-bit version of FORM was thoroughly tested
and the performance was compared to the the old 32-bit FORM.
The speedup is almost a factor 2 for small problems; for a large
size realistic application (the calculation of a quite complicated
3-loop Feynman diagram) the run-time dropped from 6665 seconds to 4105 
seconds by using the larger word size only. Another speedup of about 
1.5 was achieved simply by experimenting with different compiler
optimization flags.

Another important advantage of the 64-bit version is the use of 64-bit
addresses in the management of buffers and files. Using 32-bit pointers
the maximum size of files that can be handled is 2GB. In order to deal
with larger expressions one has to distribute the data to several files
or implement extended position information. This has not been done yet
in the 32-bit version of FORM. For the 64-bit version it is not necessary
anymore.

The next step before parallelization was profiling FORM in typical
applications in order to determine the functions that consume most of 
the run time. The following is the output of the Digital Unix profiling tool 
{\tt prof} when running a yet not very large FORM application program.
It shows that the CPU usage is well leveled between functions involved in 
the generating (marked with a G) and in the sorting of terms (S). 

\begin{verbatim}
            %time     seconds  cum %   cum sec  procedure

             20.2    501.6201   20.2    501.62 Normalize             G
             13.4    331.3018   33.6    832.92 Compare               S
              8.5    210.6748   42.1   1043.60 PutOut                S
              5.7    141.9512   47.8   1185.55 StoreTerm             S
              3.4     83.7432   51.2   1269.29 TestSub               G
              3.0     73.2539   54.1   1342.54 MergePatches          S
              2.9     72.4668   57.0   1415.01 InsertTerm            G
              0.9     21.7793   87.5   2171.72 EndSort               S
\end{verbatim}

\noindent
The time needed for compiling the program text into the FORM-internal 
representation is not dependent on the size of the problem. In realistic 
applications it is usually negligible.
 

\newpage
\noindent
{\large\bf 2. Evaluation of Different Parallel Platforms}
\vspace{5mm}

\noindent
During the part of our project that is presented here the following 
hardware has been used:

\begin{itemize}
\item
Digital workstation cluster (at TTP Karlsruhe) running DEC UNIX 4.0D\\
8 nodes with 600MHz Alpha 21164A (ev56) processors and 512MB RAM.

\item 
PC cluster (at TTP Karlsruhe) running Linux 2.2.13\\
4 nodes with 500MHz Intel Pentium III processors and 256MB RAM.

\item
IBM SP2 (at RZ Universit\"at Karlsruhe) running AIX 4.2.1\\
160 thin P2SC nodes with 120MHz processors and 512MB RAM
(256 nodes in total).
\end{itemize}

\noindent
For the implementation of the parallel version of FORM the message passing
libraries MPI, mostly the MPICH implementation (version 1.1.12) \cite{mpich},
and PVM (version 3.3.11) \cite{pvm} were used in order to guarantee maximum
portability. At the moment PVM is still an important alternative to MPI, but
we expect PVM to vanish over the years as several UNIX vendors have already 
announced to discontinue further development of their PVM libraries. Both 
message passing systems can make use of specialized device drivers underneath
in order to yield maximum communication performance.

\vspace{5mm}
\noindent
{\large\bf 2.1 Network Hardware}
\vspace{5mm}


\noindent
In the case of the DEC Alpha cluster a variety of network hardware and
various implementations of the IP (Internet Protocol) and the device
drivers for the message passing libraries exist and have been used:

\begin{itemize}

\item DEC DE500-BA Fast Ethernet NIC (100 MBit/s), 
12 port 3COM SuperStack 3000 switch,\\ DEC-UNIX TCP/IP, MPICH/P4.
\item DEC DEGPA-SA Gigabit Ethernet NIC (1000 MBit/s), 
no switch (two nodes only),\\ DEC-UNIX TCP/IP, MPICH/P4.
\item Myricom Myrinet-SAN NIC 32-bit 33MHz (1.26 GBit/s)\cite{myricom},
dual 8 port Myrinet SAN switch,\\ DEC-UNIX TCP/Myrinet IP, MPICH/P4,
\item Myricom Myrinet-SAN NIC 32-bit 33MHz (1.26 GBit/s), 
dual 8 port Myrinet SAN switch,\\ ParaStationII\cite{parastation2}
software, MPICH/PSM.
\end{itemize}

\noindent
Also for the Intel Pentium III cluster two different combinations of 
network hard-- and software have been used:

\begin{itemize}
\item 3COM 3C905B Fast Ethernet NIC (100 MBit/s), 
12 port 3COM SuperStack 3300 switch,\\ LINUX TCP/IP, MPICH/P4.
\item Myricom Myrinet-SAN NIC 32-bit 33MHz (1.26 GBit/s)\cite{myricom},
dual 8 port Myrinet SAN switch,\\ Myricom GM software, MPICH/GM.
\end{itemize}

\noindent
For a thorough understanding of the parallel systems' behavior it is of 
course crucial to compare their communication performance. For the IP 
drivers the TCP (Transfer Control Protocol) performance was determined 
(on workstation clusters the standard low-level device P4 of MPICH uses 
TCP/IP for the communication over networks). 
Moreover we measured the bandwidth and latency performance for the MPI(CH)
and PVM libraries under different conditions. Here we present the results
for the TCP and MPI(CH) measurements only. They are most important for our 
purposes.


\newpage
\noindent
{\large\bf 2.2 TCP Performance}
\vspace{5mm}

\begin{figure}[h!]
\begin{center}
\epsfig{file=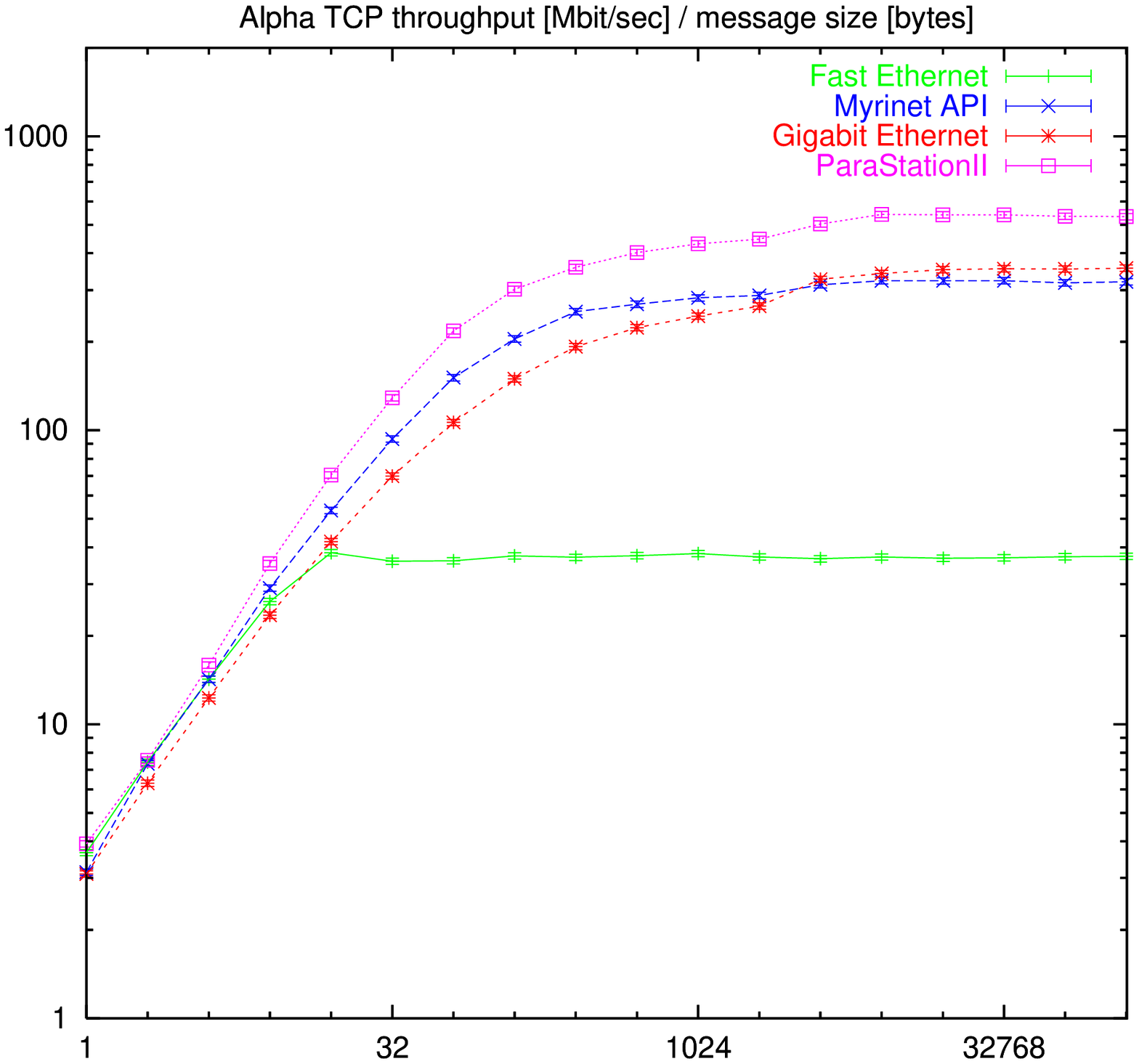,height=7.5cm,angle=-90}
\epsfig{file=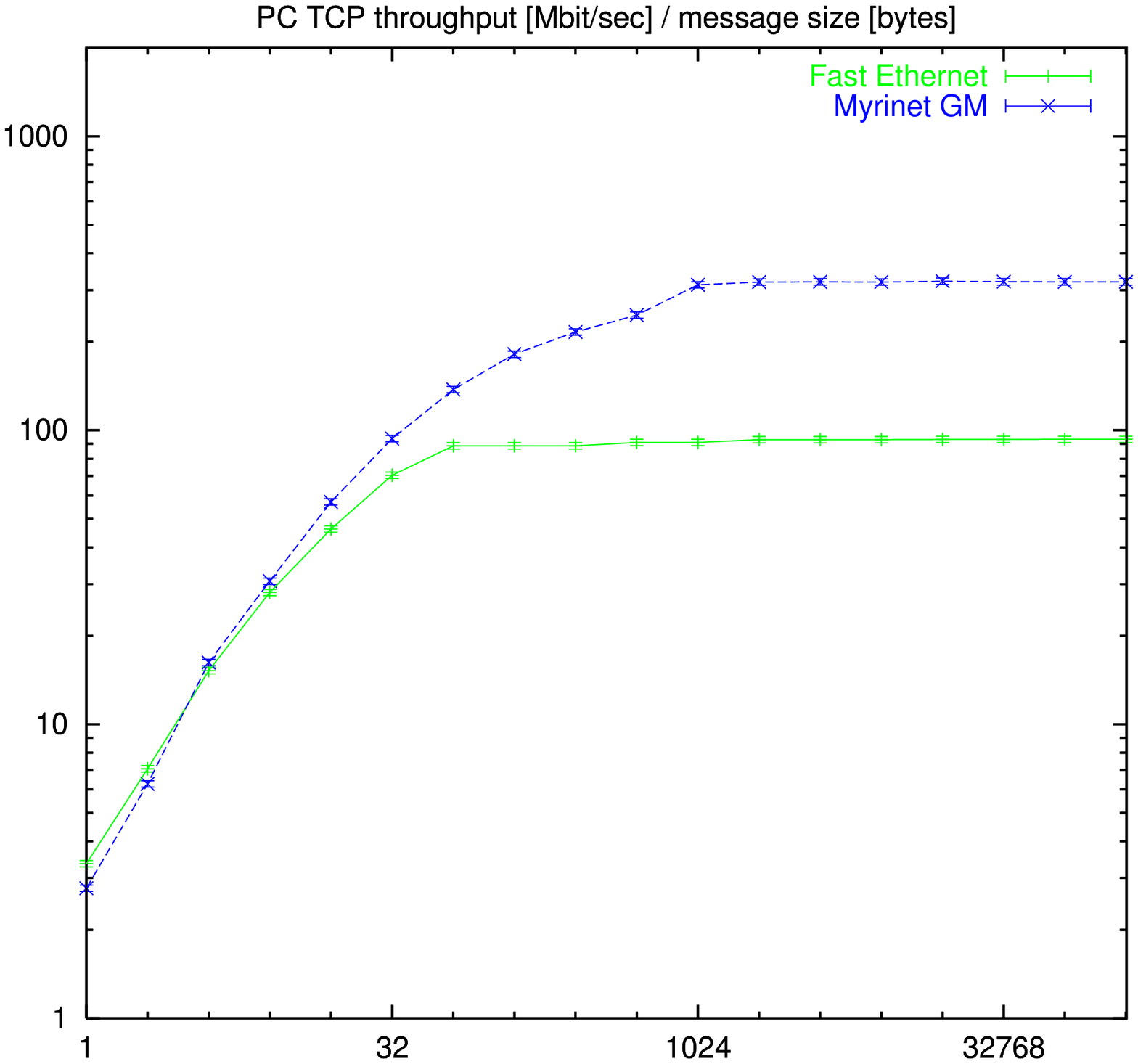,height=7.5cm,angle=-90}
\end{center}
\caption{TCP bandwidth [MBit/s]/message packet size[bytes] of different IP 
drivers on the DEC Alpha (left) and the PC cluster (right).}
\label{decpctcpperf}
\end{figure}

\noindent
The TCP stream (data transfer) performance of the IP drivers for the DEC Fast
Ethernet, the DEC Gigabit Ethernet, the Myricom Myrinet and the ParaStationII
on the DEC Alpha cluster and for the 3COM Fast Ethernet and the Myricom
Myrinet for the Intel Pentium III PC cluster was measured with the standard
package netperf~\cite{netperf}. It provides a script that repeats the 
measurement until a certain level of accuracy is reached. A couple of 
parameters can be chosen, namely the buffer sizes on both the sending \&
receiving side and the size of the transmitted packages. The results are
shown in figure \ref{decpctcpperf} for optimum buffer sizes for each
architecture.

\noindent
The TCP throughput increases linearly for small message packet sizes. 
Obviously, in this region the communication is dominated by the protocol
overhead. For large messages sizes a maximum throughput value is 
reached. For the Fast Ethernet on the Alpha cluster (left) this is about
40MBit/s only, whereas the Fast Ethernet on the PC Cluster yields a 
maximum throughput of more than 90MBit/s, which is about the expected maximum
of 100MBit/s. The maximum TCP/IP throughput of the Gigabit Ethernet on the
Alpha cluster is 350 MBit/s and for the Myrinet on both clusters a maximum
bandwidth of 320Mbit/s is achieved. This is far below the network hardware 
limit of 1Gbit/s and 1.28Gbit/s resp. The ParaStationII TCP does not use
an IP protocol underneath and therefore gives a better maximum performance
of 530Mbit/s. The results for ParaStationII and Myrinet GM are shown for 
completeness only. The corresponding MPI implementations do not use TCP/IP.


\vspace{5mm}
\noindent
{\large\bf 2.3 MPI Performance}
\vspace{5mm}


\noindent
The performance of the MPI(CH) libraries was examined using the Pallas
MPI benchmarks (PMB) \cite{pmb}. Again a script is provided that can be
used to measure the throughput and latency of the messages passing 
operations without further interaction. Here we present the results
for the basic ping-pong-benchmark only. It is the simplest possible single
transfer benchmark, based on the blocking MPI routines {\tt MPI\_send}
and {\tt MPI\_recv}: one process(or) sends a message of $n$ bytes to another 
process(or), which immediately sends that message back. There is no 
concurrency with other message passing activity during this test. Thus
the bandwidth and latency are measured under optimum conditions.

\begin{figure}[h!]
\begin{center}
\epsfig{file=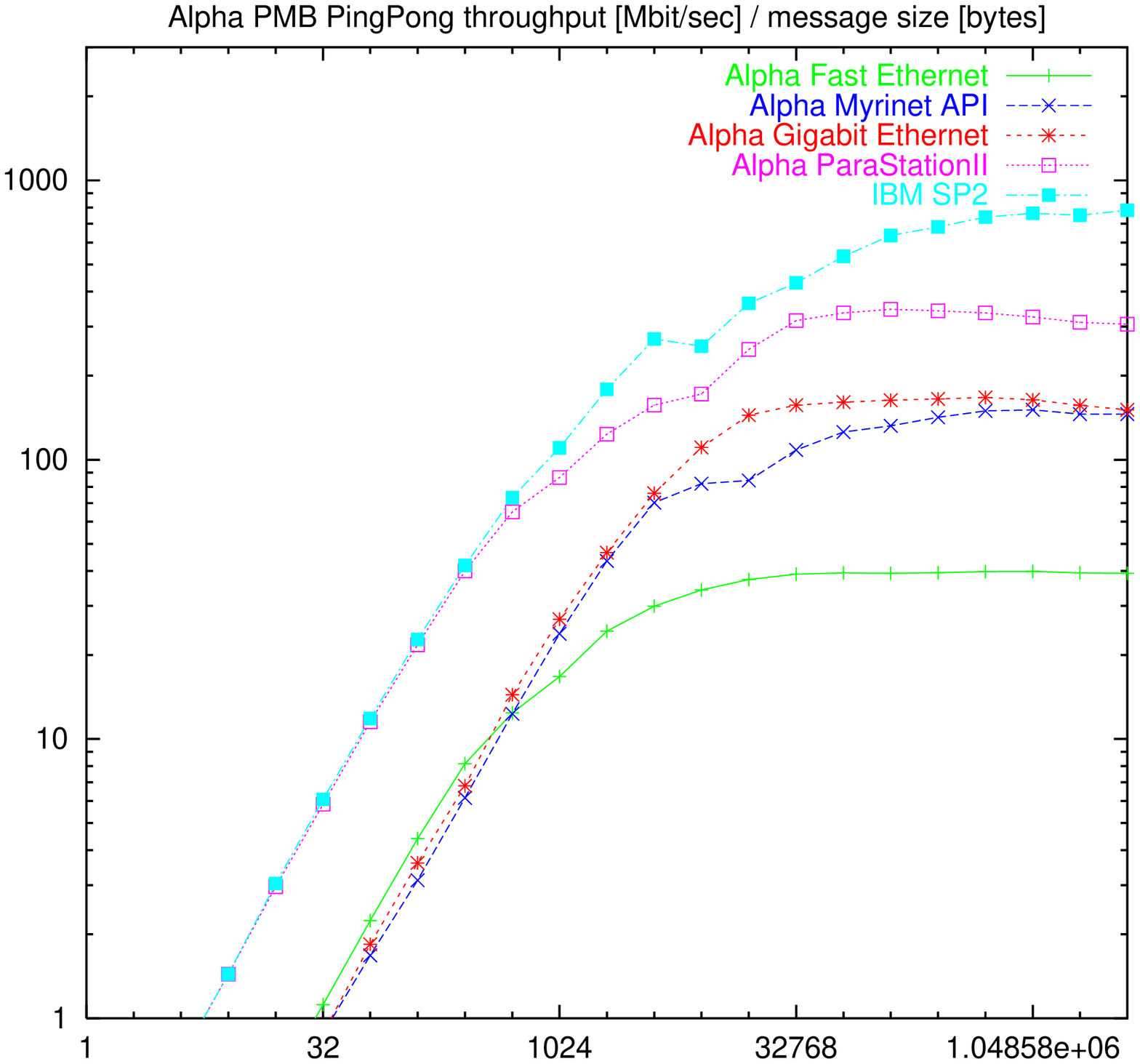,height=7.5cm,angle=-90}
\epsfig{file=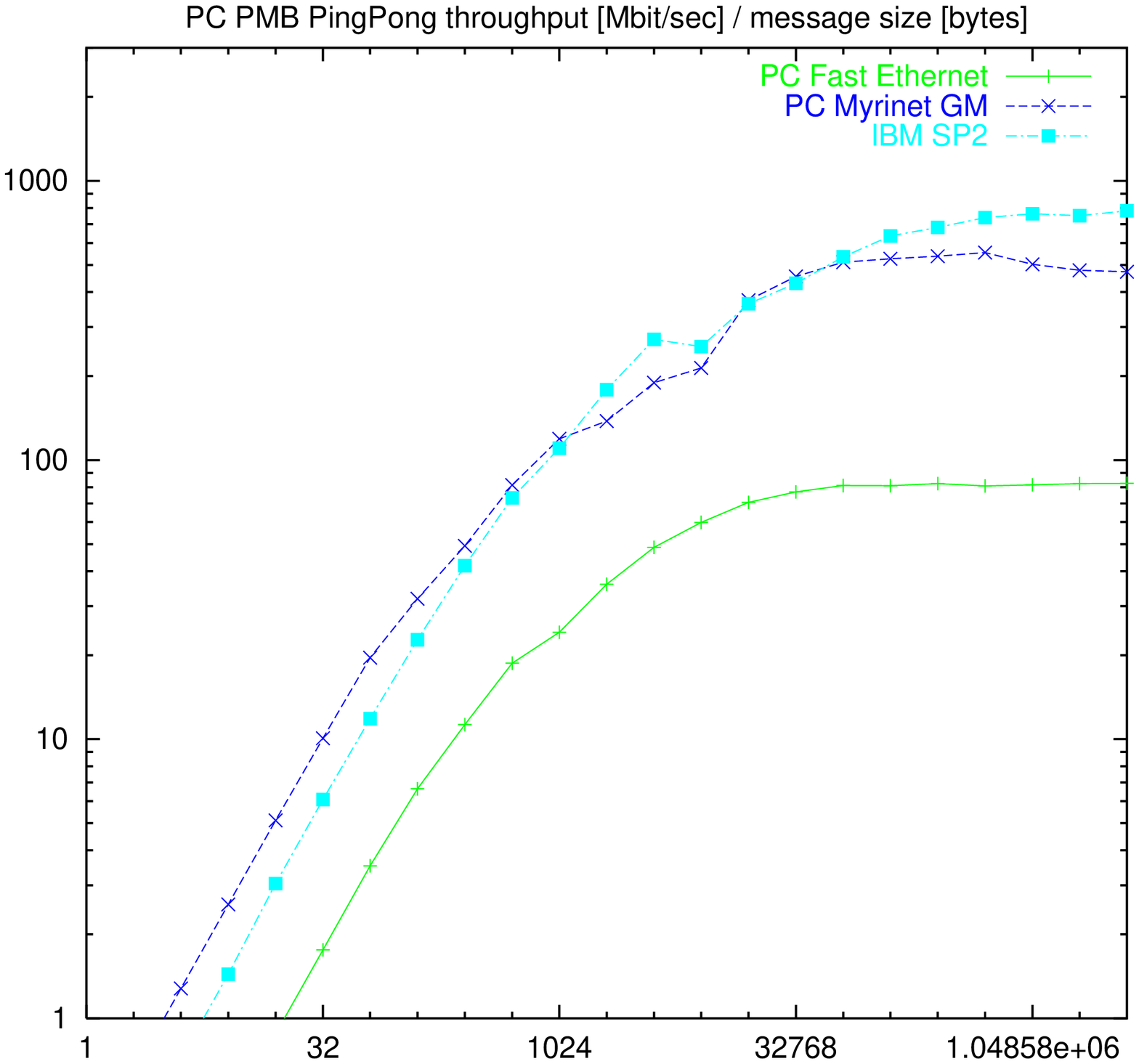,height=7.5cm,angle=-90}
\end{center}
\caption{MPI(CH) bandwidth results for the PingPong benchmark.
(left: Alpha cluster, right: PC cluster).}
\label{pinpongbw}
\end{figure}

\noindent
As can be seen from figure \ref{pinpongbw} the bandwidth increases as the 
message size does and reaches a maximum value for large message sizes just
like the TCP throughput. Obviously the performance of the different IP 
drivers directly translates to the MPI(CH) performance. On both clusters 
adding the MPICH layer causes no dramatic effect for the Fast Ethernet,
but the bandwidth for the Myrinet and the Gigabit Ethernet drops below
200Mbit/s. The implementations of MPI that do not use TCP/IP show better
performance: the maximum bandwidth of MPICH/PSM on the Alpha cluster is
350Mbit/s and of MPICH/GM on the PC cluster even 550MBit/s. The MPI 
implementation on the IBM SP2 reaches a maximum bandwidth of 800 MBit/s.

\begin{figure}[h!]
\begin{center}
\epsfig{file=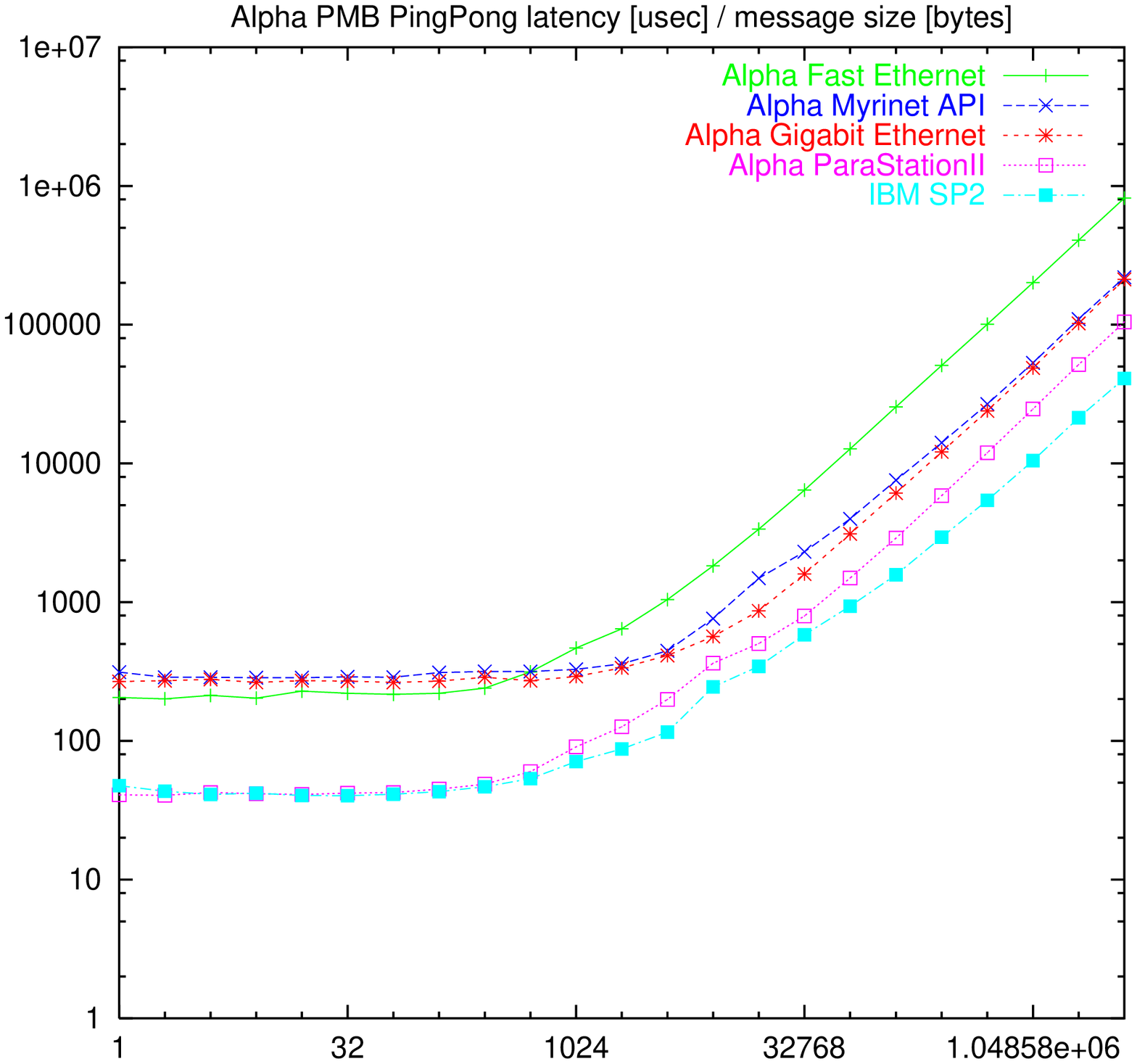,height=7.5cm,angle=-90}
\epsfig{file=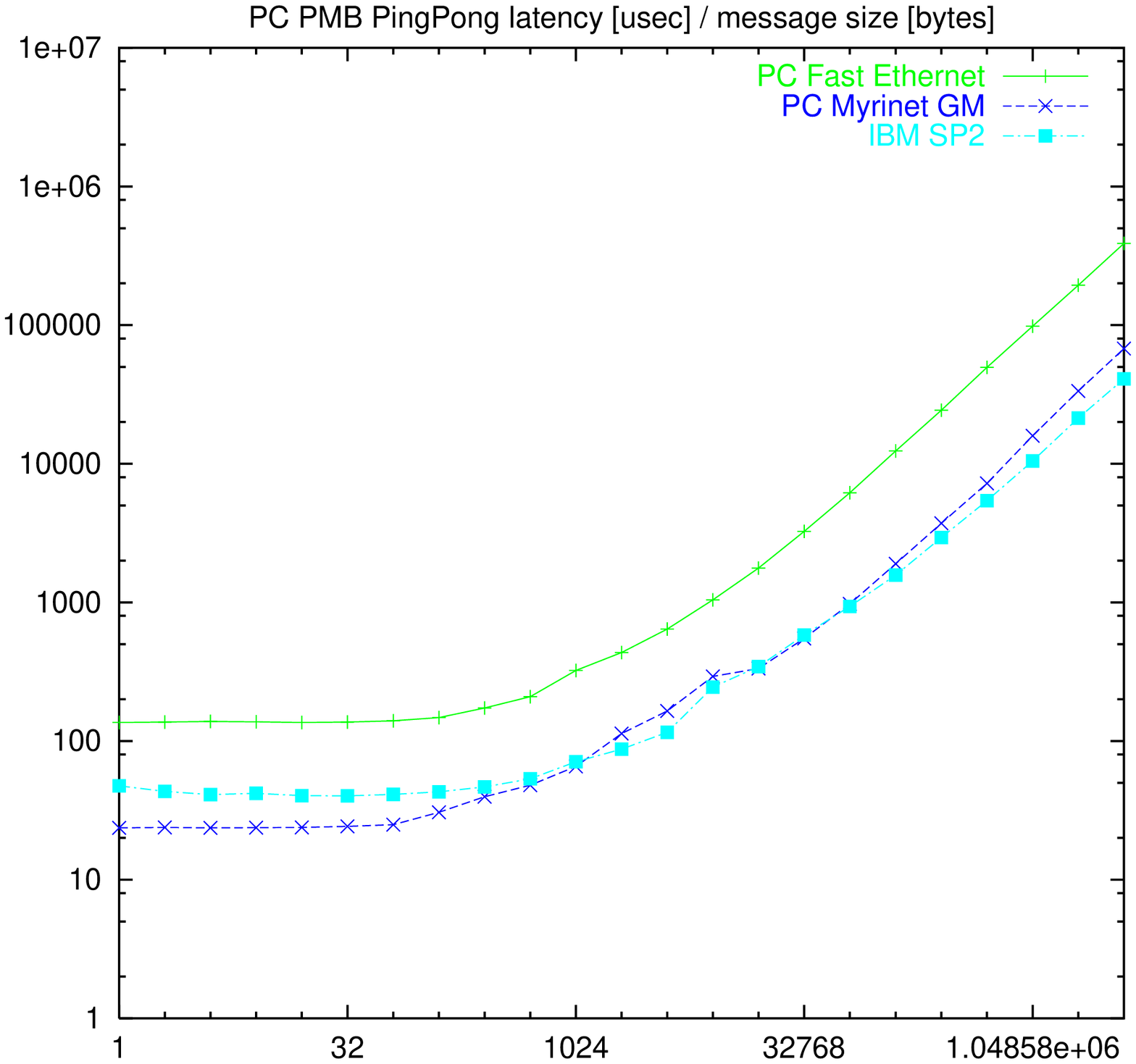,height=7.5cm,angle=-90}
\end{center}
\caption{MPI(CH) latency results for the PingPong benchmark
(left: Alpha cluster, right: PC cluster).}
\label{pingponglt}
\end{figure}

\noindent
In figure \ref{pingponglt} the measured MPI(CH) latency for the ping-pong 
benchmark is shown. The latency ranges from a few microseconds for small
messages sizes up to a second for large message sizes. Using MPI(CH)
over the IP protocol results in a minimum latency of more than 200 
microseconds. Lower latencies can be achieved with special device drivers
only. It turns out that the minimum latency for MPICH/PSM on the Alpha
cluster (about 40 microseconds) is about the minimal latency on the IBM 
SP2, but the minimum latency on the PC cluster using MPICH/GM is even 
smaller: about 20 microseconds. 

\noindent
It is important to note that the latency does not increase dramatically
up to 1024 bytes. In practical programming that means that whenever 
a small amount of data (much less than 1024 bytes) is sent one can use 
the corresponding message to transmit additional information (e.g. 
statistics) without increasing the latency. 


\newpage
\noindent
{\large\bf 3. The Parallel Version of FORM}
\vspace{5mm}

\noindent
{\large\bf 3.1 The Parallelization Concept for FORM}
\vspace{5mm}


\noindent
As has been already mentioned, FORM is specialized in working with very 
large expressions, typical jobs run for weeks and can need Gigabytes of
temporary disk space. So the idea of parallelizing FORM is obvious and 
interesting not only because distributing the CPU usage would speed the
program up but also because a distribution of the large temporary data
could make even larger problems accessible.

For running typical multi-loop calculations, where several hundred or
thousands of single diagrams have to be calculated, a ``trivial''
parallelization by distributing the diagrams to different computers
(e.g. with a distributed make-command) is an obvious solution. Still 
there is a strong interest in speeding up the calculation of complicated
single diagrams in both the programming and debugging phase of a project
and production jobs.

The limitation of performing almost only local operations makes FORM very 
well suited for parallelization and the concept of parallelization is 
straightforward: 
distribute the input terms among the available processors, let each of them 
perform the local operations on its input terms and generate and sort the 
arising output terms. 
At the end of a module the sorted streams of terms from all processors have 
to be  merged to one final output stream again. 
The compiling of the program-text to the internal representation was 
considered to not be worth parallelizing. 
This concept indicates to use a master-slave structure for the 
parallelization, where the master would store the expressions and distribute
and recollect all the terms of each expression.

For the implementation of this raw concept we used a four step strategy:
\begin{itemize}
\item one process(or) generates terms, a second process(or) sorts the output 
terms
\item instead of only one process arbitrary many processes perform the sorting.
\item the input terms are distributed and the term generation is also done in
parallel
\item full FORM functionality, avoid or handle worst cases, load-leveling, fault tolerance
\end{itemize}
This approach has several advantages, the most important being that having 
working versions in every stage gives us a good idea of how good the 
parallelization is and the possibility of realistic tests even at a very 
early stage.


\vspace{5mm}
\noindent
{\large\bf 3.2 The Two-Processor Version}
\vspace{5mm}


\noindent
This first step of the parallelization turned out to be very useful, since
it not only gave a deep insight of how changes to the source code of FORM 
have to be made without affecting the efficiency of the well optimized 
sequential code. It also served as a check of whether and how the concept 
could or could not lead to a decent speedup. 
It could be shown that for parallelizing software on a cluster of very fast 
workstations the importance of avoiding communication overhead can not be 
overstated, especially concerning the latency of network-communications.
It was clear after these experiments that all the communication had to be
done in a buffered way, since sending around single terms increased the run
times of the two-processor code up to a factor 20. With the buffered version
the run-time could be limited to about 1.5 times of that of the sequential 
code with two workstation connected by a 10 Mbit/s Ethernet using the PVM 
and MPI(CH) libraries. This was considered to be fast enough since in this
case due to only minimal work-load overlap we merely were adding the 
communication overhead.
        

\vspace{5mm}
\noindent
{\large\bf 3.3 Parallel Sorting}
\vspace{5mm}

\begin{figure}[h!]
\begin{center}
\epsfig{file=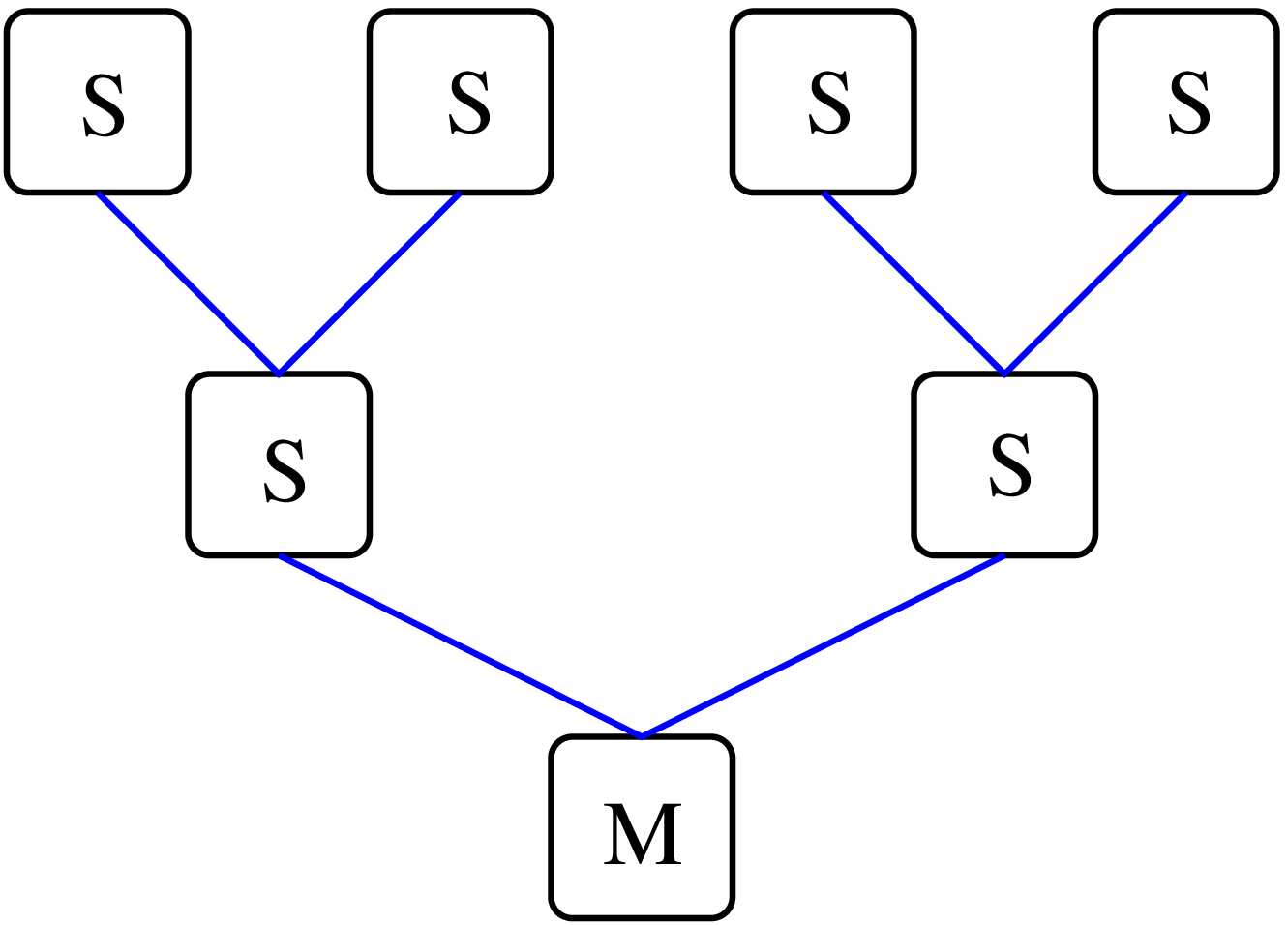,height=3.8cm,width=5cm}
\hspace{2cm}
\epsfig{file=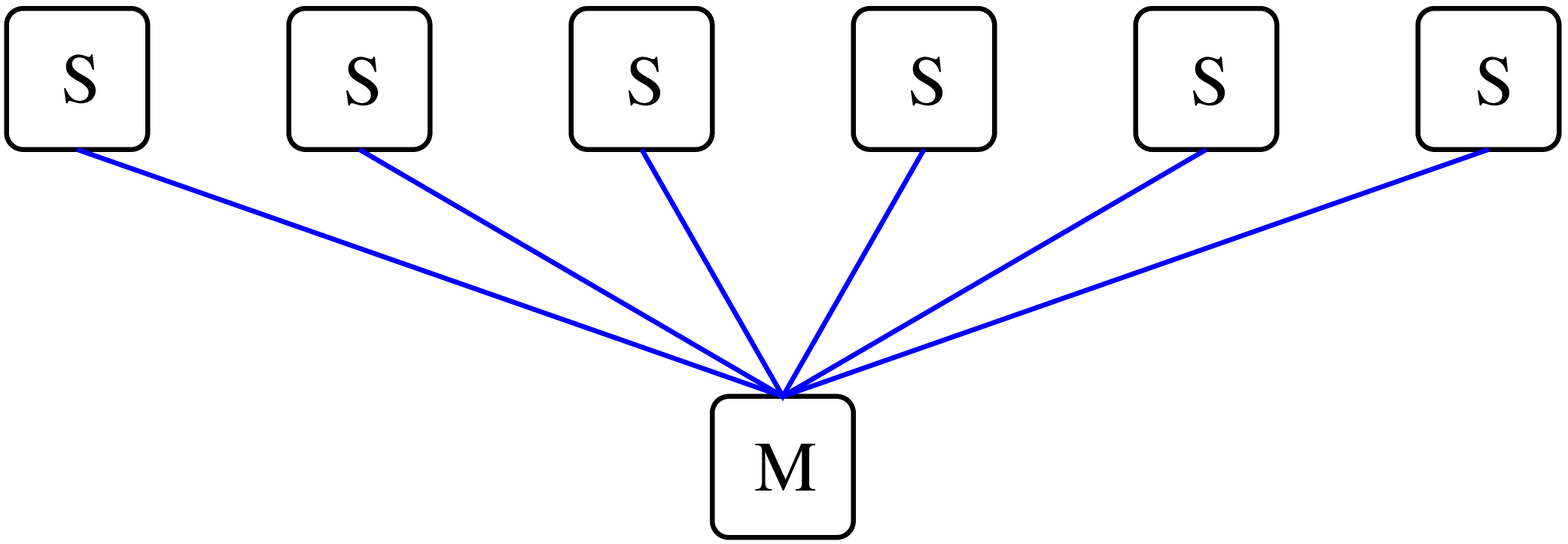,height=2.7cm,width=7cm}
\end{center}
\caption{Two trees for merging terms from different processors}
\label{sorttrees}
\end{figure}

\noindent
The second step was to distribute the output terms among arbitrary many 
processors and do the sorting in parallel. Since this part of the sorting 
relies strongly on communication between the processors, it most probably
sets the limits of parallel speedup. 
A first try was to map the ``tree of losers'' used in FORM to merge sorted 
patches onto the processors as shown in figure \ref{sorttrees} (left).
While it would distribute the workload in an optimal way this approach
adds too much communication overhead. 

\noindent
This is why in the end a much simpler communication structure was used 
(see figure \ref{sorttrees}, right), where all the slaves send their 
sorted terms to the master process and this process uses a local 
``tree of losers'' to merge the output streams of the slave-processes. 
Additional effort was made to overlap the work on the master process with the 
sorting done on the slaves, which caused a much deeper interference with the 
sequential code.

\begin{figure}[h!]
\begin{center}
\epsfig{file=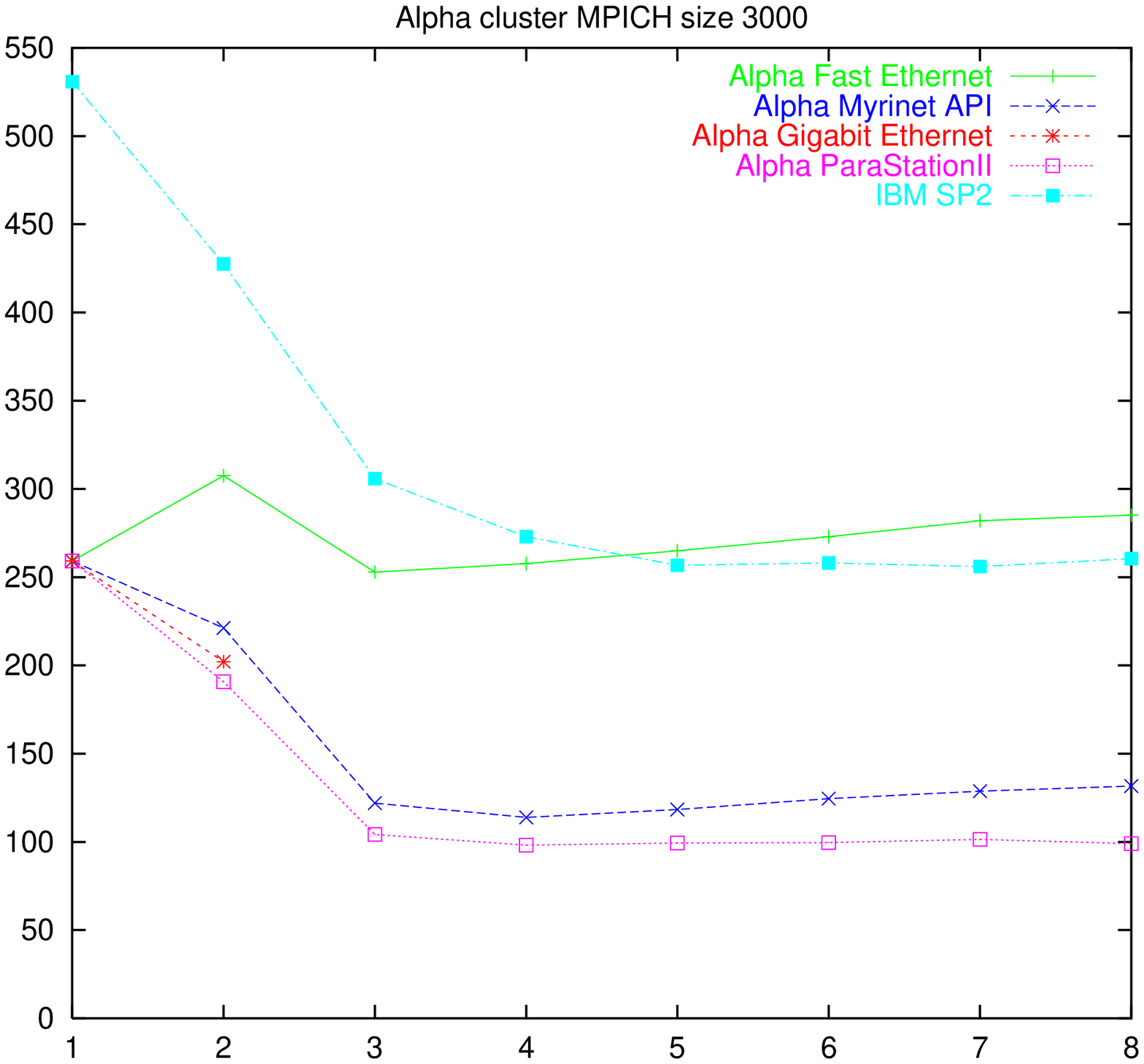,height=7.5cm,angle=-90}
\epsfig{file=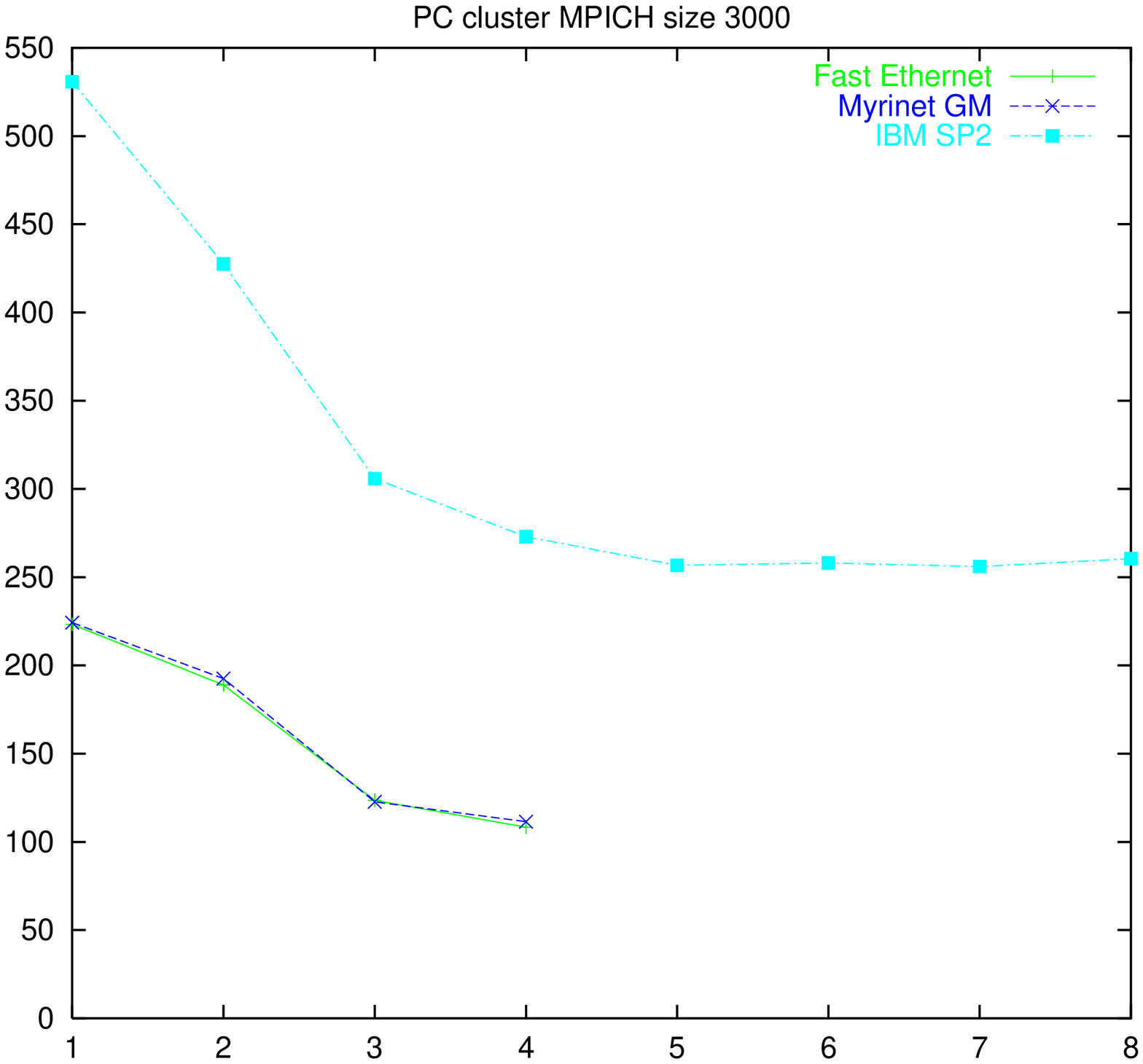,height=7.5cm,angle=-90}\\
\epsfig{file=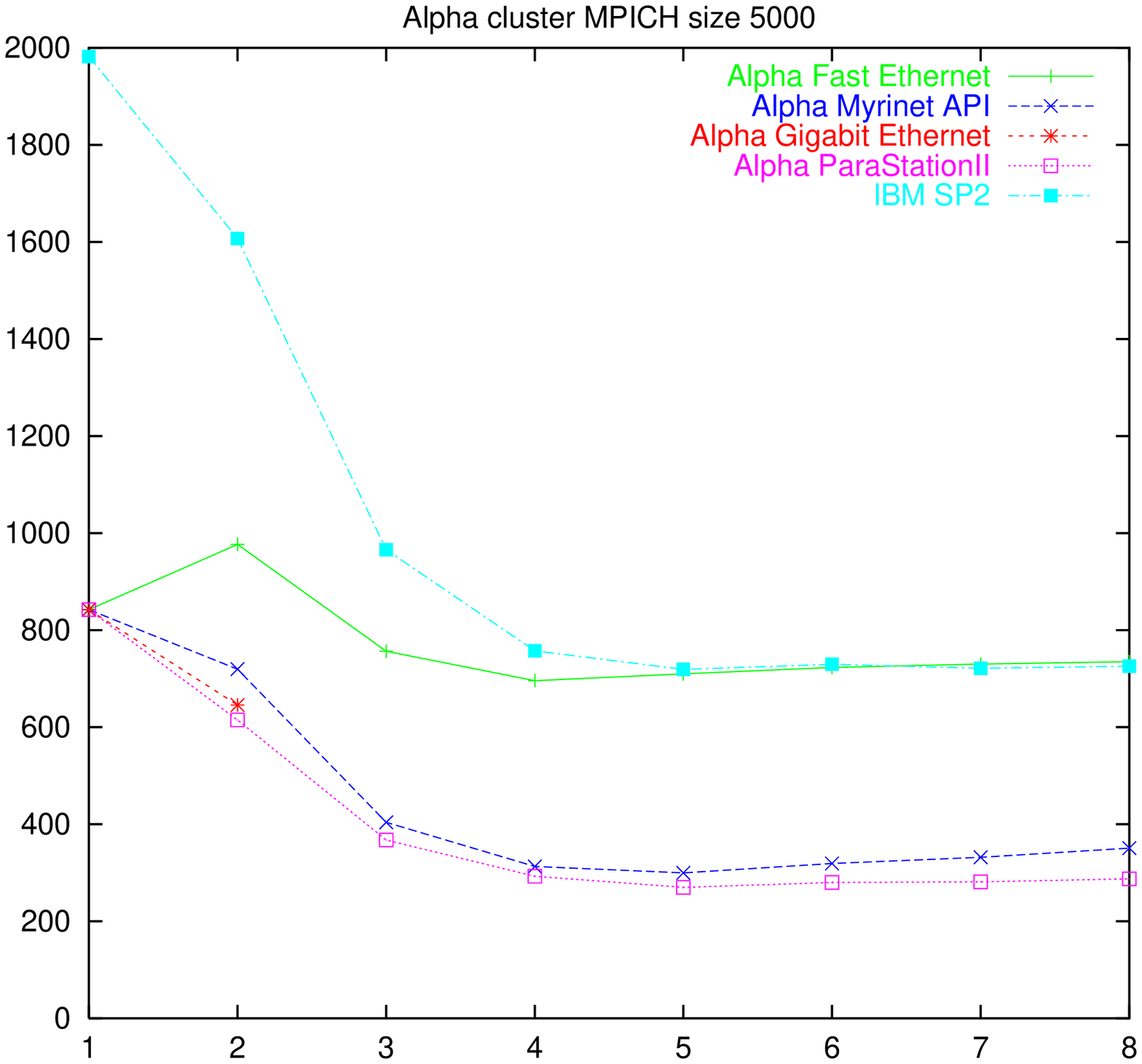,height=7.5cm,angle=-90}
\epsfig{file=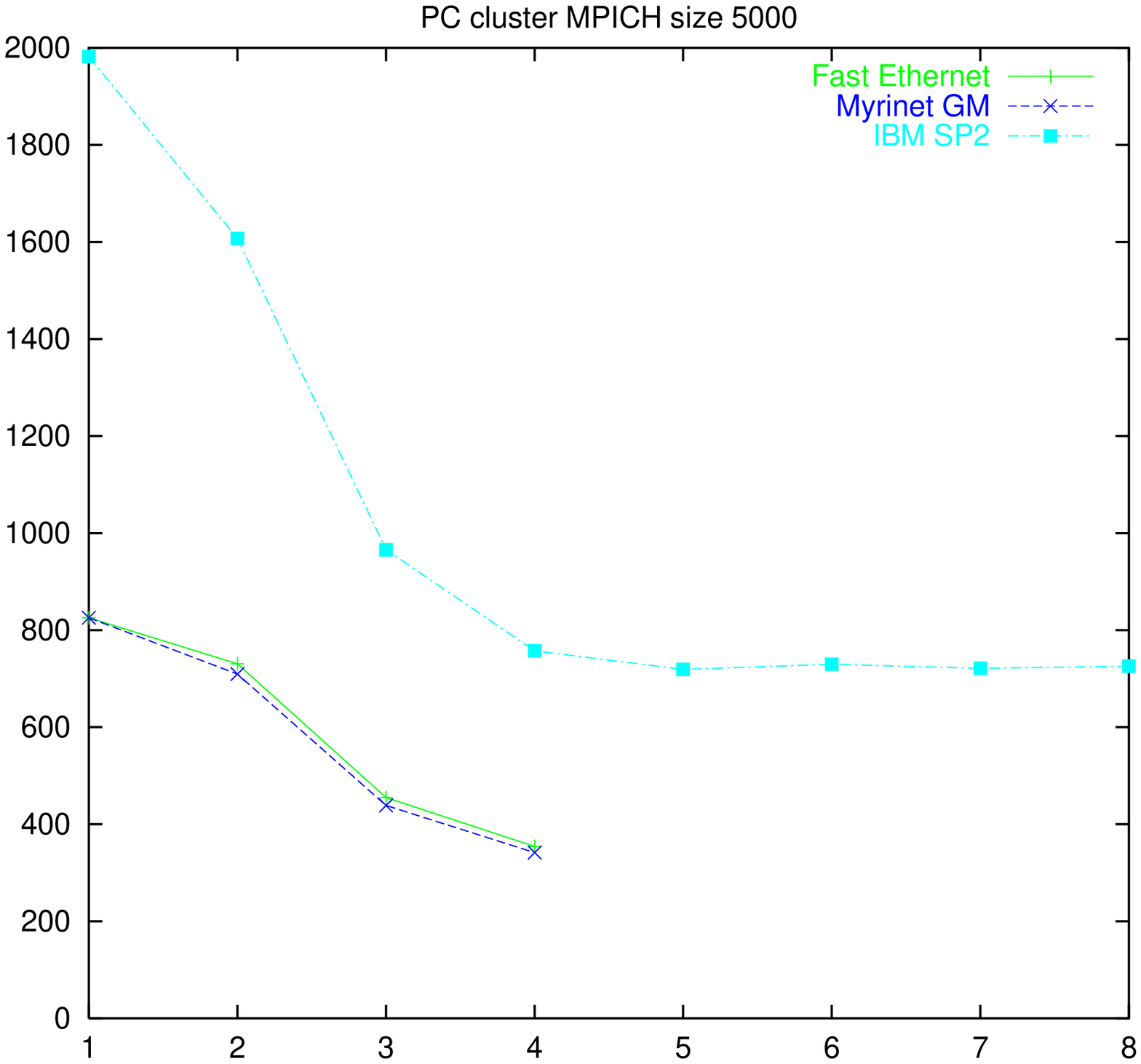,height=7.5cm,angle=-90}
\end{center}
\caption{Parallel sorting on the DEC Alpha cluster (left) and
on the PC cluster (right). The results (wall clock run-times [sec] /
number of processors) for two different problem sizes n=3000 (top) and 
n=5000 (bottom) are shown. For both architectures different hard-- and 
software combinations are compared and the results for the IBM SP2 are
shown for comparison.}
\label{sortplots}
\end{figure}

\noindent
Since only the sorting is of interest any problem that produces a sufficient
number of intermediate terms could have been used to test the parallel 
speedup. We chose a very simple program, that expands the expression 
$(a_1+\cdots+a_n)^2$ and then replaces $a_1$ by $-(a_4+\cdots+a_n)$ which 
results in a short, easy to check result and---by choosing different values
of $n$---can be scaled in an easy way. 
The run-times we measured with this version and different
combinations of communication hard-/software are shown in figure 
\ref{sortplots} for two different values of $n$: $n=3000$ (top) and 
$n=5000$ (bottom).
The corresponding problem sizes are $\sim 400$MB and $\sim 1400$MB resp.~for
the 64-bit architecture of the Alphas and roughly half as large for the \
32-bit systems. 
We find that the performance of the Fast Ethernet on the Alpha cluster is 
so low that no speedup can be achieved. All other systems yield a maximum 
speedup of about a factor of 2.5 that is apparently independent of the 
problem size. Obviously the generating processor is able to keep about 4 
processors busy in sorting only, i.e. there is a rather quick saturation in
the speedup.
As a result the runtimes of the parallel sorting version of FORM on the IBM
SP2 hardly reach the runtimes of the sequential version on the Alpha and the
PC clusters. On the other hand the speedup is not demolished again by the
use of too many processors, a fact that has been explicitly checked with
32 processors on the IBM SP2. 
 

\vspace{5mm}
\noindent
{\large\bf 3.3 Parallel Generating}
\vspace{5mm}


\begin{figure}[h!]
\parbox{0.495\textwidth}{
\epsfig{file=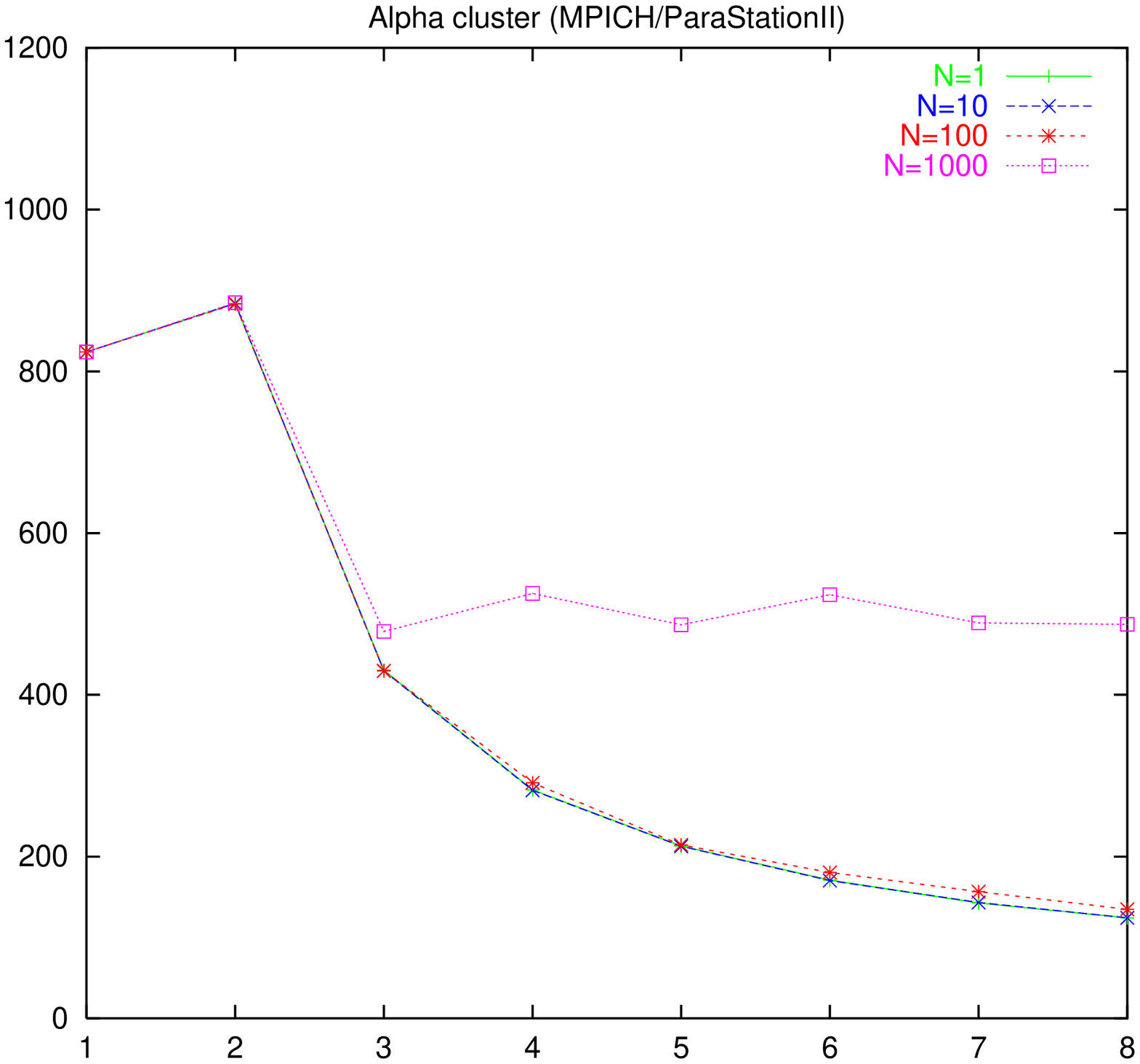,height=7.5cm,angle=-90}}
\parbox{0.495\textwidth}{
\epsfig{file=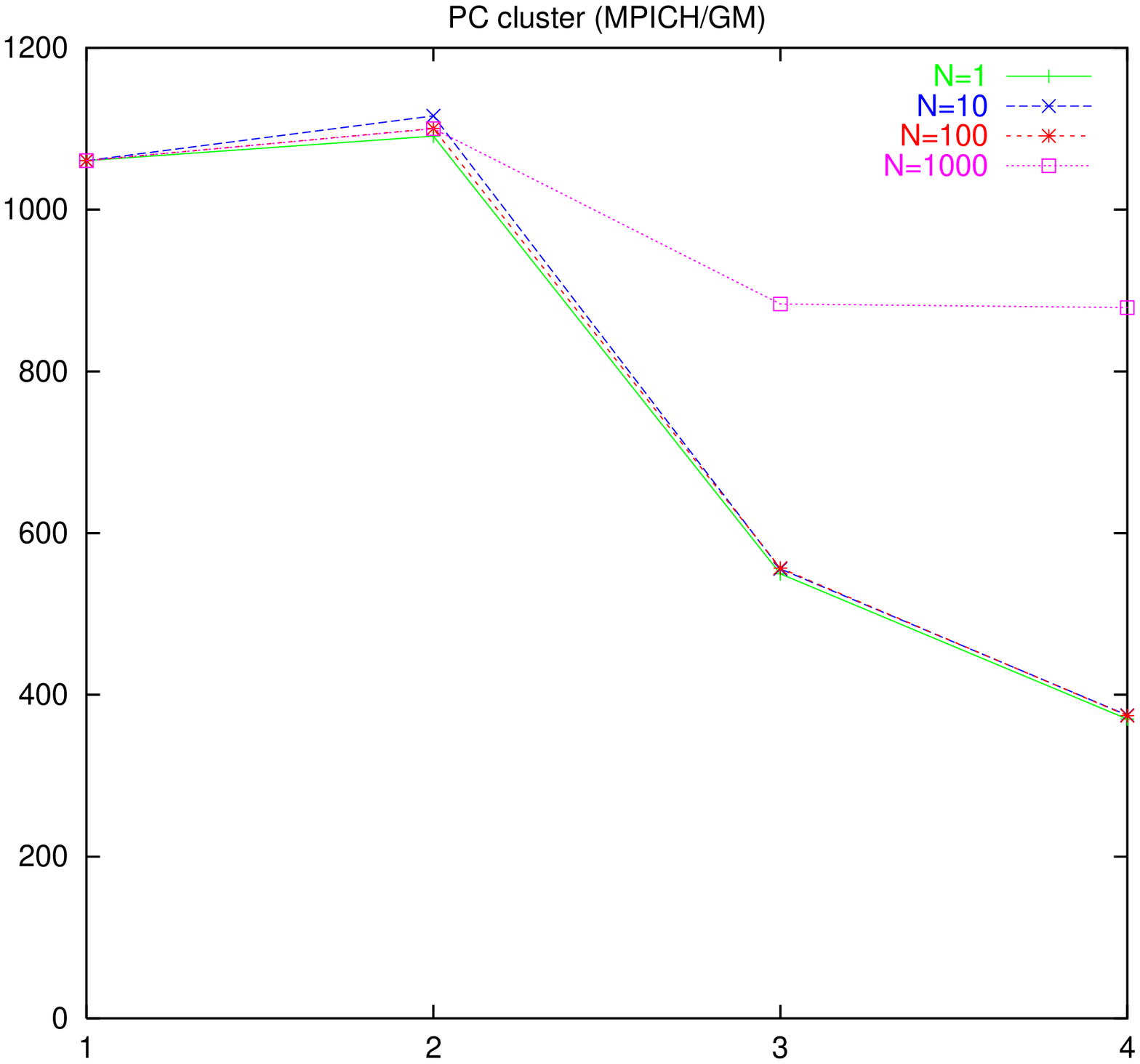,height=7.5cm,angle=-90}}
\parbox{0.495\textwidth}{
\epsfig{file=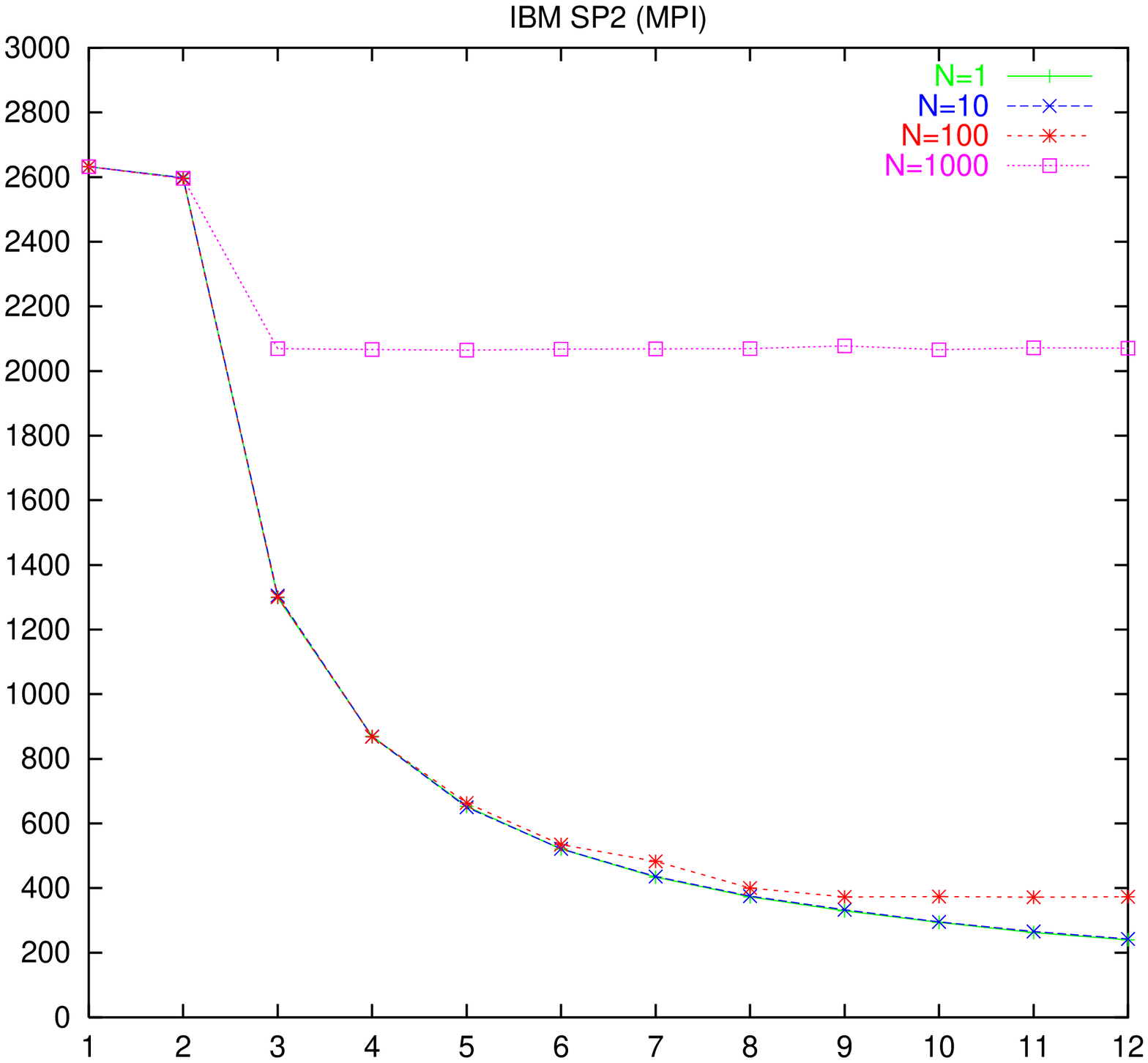,height=7.5cm,angle=-90}}
\rule{0.02\textwidth}{0pt}
\begin{minipage}{0.4\textwidth}
\caption{Results for the parallel generating version running a test
program that amounts to ideal input for the parallel version. Shown
are the run-times[sec]/number of processors for the different 
architectures: the Alpha cluster (upper left), the PC cluster (upper
right) and the IBM SP2 (left).}
\label{idealinputnobrack}
\end{minipage}
\end{figure}

\noindent
The last step towards a working prototype is the distribution of input 
terms among all processes before generation of the output terms. 
The slave-processes of course also need all the information necessary 
for the generation of the terms, which is at the moment realized by having 
them all read the program text file and compile their own internal 
representation and broadcasting only the rest of necessary information 
from the master process. 

\noindent
Just as with the sorting the speedup is strongly dependent on the number of 
terms sent within one message. In the current implementation this number can
be changed at the start of the program, but it might be a good idea to 
adjust this number module-wise.
Of course choosing a too coarse grained distribution will result in the 
danger of running into worst cases, where all the work sits in only one of 
the input patches, and only one processor is busy.
On the other hand, the fine grain distribution causes more overhead. 
The best setting turned out to be not only dependent on the underlying 
soft/hardware, but also strongly dependent on the problem that is run. 
The distribution is organized such that the master sends a patch of terms
to each slave process at the beginning of the module and then waits for the 
slaves to ask for new terms whenever they are finished with their last patch 
of input terms. This actually turns the concept in that of a client-server
situation which will also be useful to make the slaves receive any kind 
of global information when the need arises. 
It also produces a decent load leveling among the slaves, which can be 
controlled by the size of the input-patches and could even be adjusted
during run-time for further improvement.

\noindent
Figure \ref{idealinputnobrack} shows the results for a test program that 
amounts to ideal input for the parallel generating version. It starts out 
with a bit more than 2000 terms. On each of these terms a lot of work is
done and a huge number of intermediate terms is produced. In the end the 
result collapses to a rather small expression.

It is obvious that due to the client-server-like parallelization strategy
one processor (the server-node) is kept busy with the organization of the 
parallel data flow and a second processor alone does hardly yield a speedup.
With this exception the speedup is found to be almost linear for all 
architectures. In contrast to the parallel sorting version the fully
parallelized version of FORM gives such a high speedup on the IBM SP2
that it becomes competitive to the Alpha and the PC cluster. As expected for
this particular test program the dependence on the number of input terms
distributed at once is not strong, because the work is uniformly distributed
among the terms. Obviously the distribution of a thousand terms at once is 
too coarse grained and results in a more or less straight line in the speedup
plots for a number of processors $\geq 3$ (i.e. more than 2 slave processes).


\vspace{5mm}
\noindent
{\large\bf 3.4 Special Solutions to Non-Local Operations}
\vspace{5mm}


\noindent
There are a few non-local operations that are particularly useful. In order
to at least partially parallelize these operations special solutions had 
to be found:

\begin{itemize}

\item preprocessor instruction {\tt \#redefine}

The preprocessor instruction {\tt \#redefine} allows to change a 
preprocessor variable depending on the occurrence of certain terms.
At first glance this seems to cause problems for the parallel version,
when different slave processes assign different values to the same
preprocessor variable. In fact it is sufficient to let the master know
the last (w.r.t.~the input terms) modified value of the preprocessor
variable. This already guarantees that the parallel version uses the
same value for the preprocessor variable in the next module that the
sequential version would have used.

\item module option {\tt polyfun}

The {\tt polyfun} module option is used to add up the arguments of certain
functions (such as coefficients). In some applications this reduces the
number of terms drastically. In the parallel version only the ``tree
of losers'' on the master-process had to be modified. The slaves use
the usual sort routines that already give the correct behavior.

\newpage
\item {\tt collect} statement

The {\tt collect} statement collects bracketed expressions in function
arguments. In the parallel version this can completely be done by the
master-process.

\item {\tt keep brackets} statement

The {\tt keep brackets} statement hides the contents of brackets from
execution of the statements of a module. The parallelization of this
feature is not straightforward and only partially implemented in the
current version. But corresponding tests show that this version is
sufficient for real applications.


\end{itemize}


\vspace{5mm}
\noindent
{\large\bf 3.5 A Real Application: Moments of Structure Functions}
\vspace{5mm}


\noindent
With the additions discussed in the last section it was eventually possible
to run the wide-spread FORM-package MINCER \cite{mincer}, which can calculate 
certain types of Feynman diagrams up to the three loop level. First some 
easy standard integrals were calculated in parallel and proven to come out 
correctly. After that the computation of diagrams of a still ongoing project,
the calculation of higher moments of structure functions 
\cite{moments} was chosen as a real application to test the parallel version
of FORM.

\begin{figure}[h!]
\parbox{0.495\textwidth}{
\epsfig{file=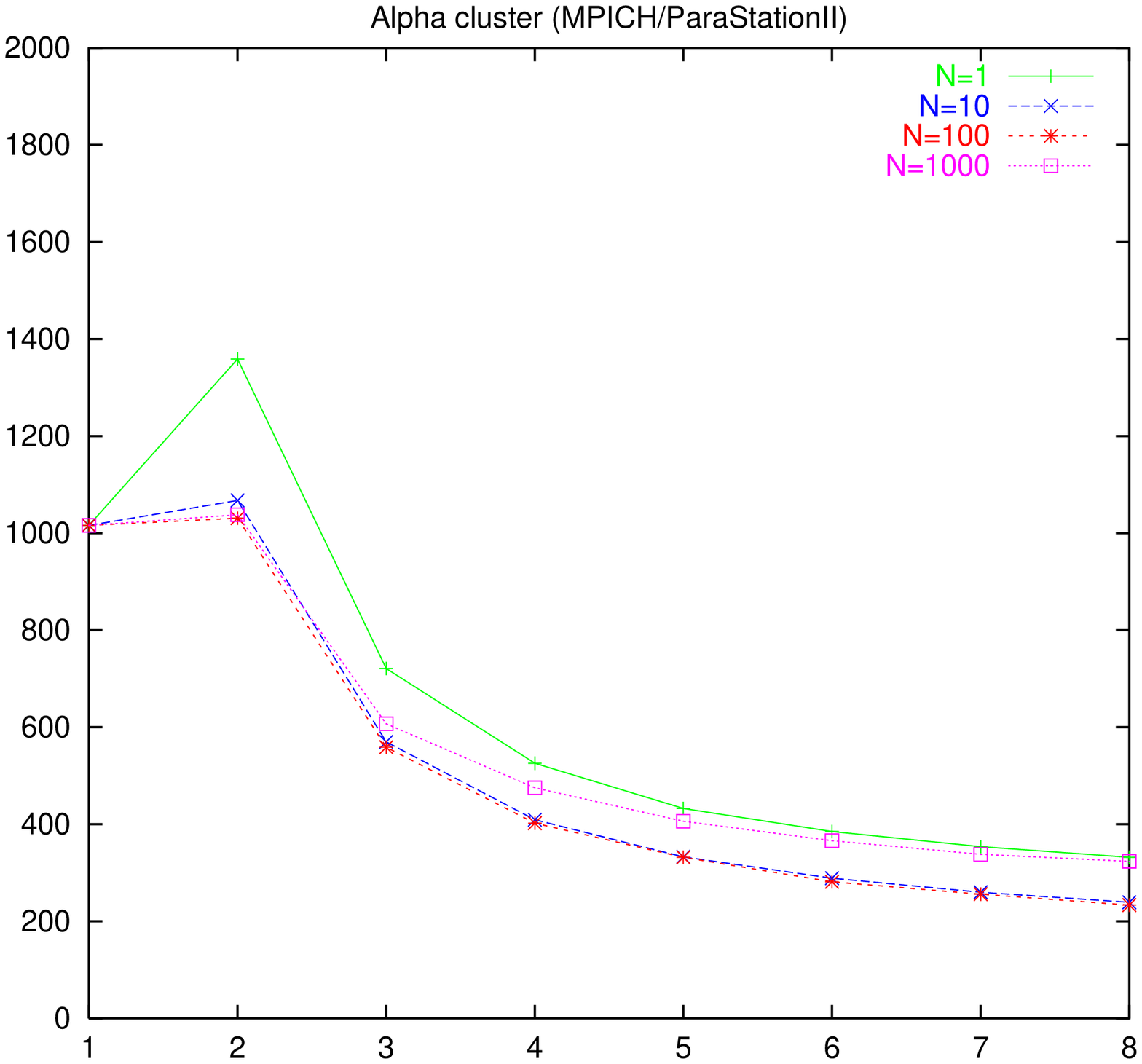,height=7.5cm,angle=-90}}
\parbox{0.495\textwidth}{
\epsfig{file=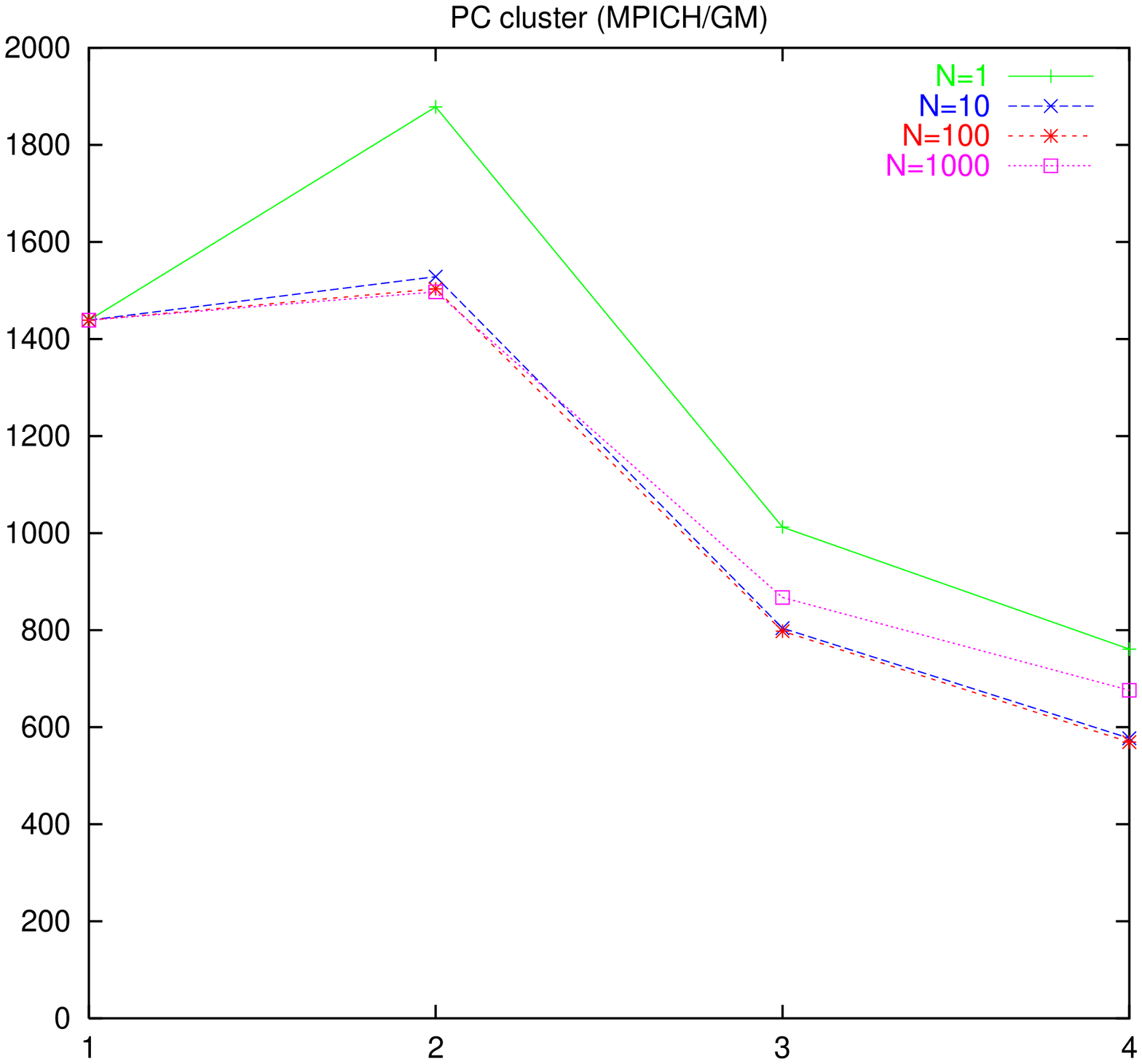,height=7.5cm,angle=-90}}
\parbox{0.495\textwidth}{
\epsfig{file=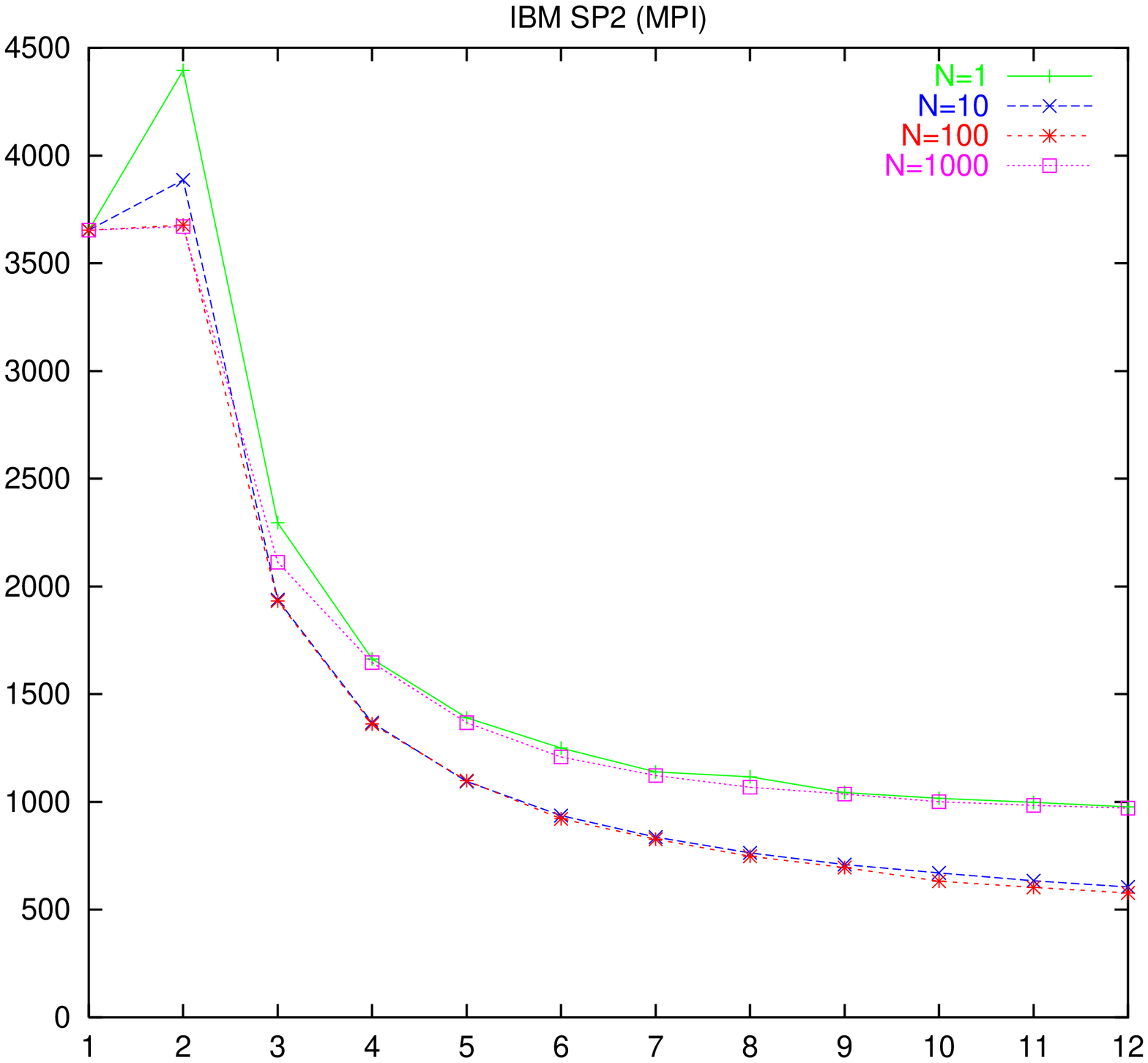,height=7.5cm,angle=-90}}
\rule{0.02\textwidth}{0pt}
\begin{minipage}{0.4\textwidth}
\caption{Results for the fully parallelized version running MINCER on
a typical 3-loop diagram of an actual calculation. Shown are the 
run-times[sec]/number of processors for the different architectures:
the Alpha cluster (upper left), the PC cluster (upper right) and the
IBM SP2 (left).}
\label{diagram1}
\end{minipage}
\end{figure}

\noindent
When running MINCER usually over a hundred modules are executed (depending
on the diagram under consideration). In most of those only very few terms
are active, which amounts to `worst cases' for the parallelization. Also 
the results are received from the FORM-code written for the sequential 
version of FORM {\it without any} modification or optimization. This 
corresponds to a perfect code reuse. Figure \ref{diagram1} shows the 
results for a typical diagram. Obviously the speedup is not linear any more,
but reasonable: a factor of 2.5 with 4 nodes on the PC cluster using MPICH/GM,
a factor of 4.5 with 8 nodes on the Alpha cluster using MPICH/PSM and a factor
of 6 with 12 nodes on the SP2. For the problem under consideration obviously
distributing single terms ($N=1$) or too large patches of terms ($N=1000$) is
not a good idea and a value in the range of $10\ldots 100$ yields best 
performance.

\vspace{5mm}
\noindent
{\large\bf 4. Conclusion \& Outlook}
\vspace{5mm}


\noindent
We have shown that at least for a symbolic manipulation program like
FORM, which with a few exceptions allows local operations only, a 
successful parallelization can be done even on workstation clusters
using message passing libraries.  
Of course the speedup that can be achieved depends strongly on the
problem under consideration. Ideal input to the parallel version and 
a problem size that is not too small yields a speedup that is linear
in the number of slave processors. For realistic complex applications
the speedup is still considerable. It is particularly noteworthy that
these results have been achieved with FORM programs that were written
for the sequential version and have not been modified. Generally the
FORM user does not have to know anything about the mechanism behind the
parallel version to run already existing programs in parallel. Still,
some knowledge can help to tune them and achieve a higher speedup.

\noindent
The performance and stability of the high performance Myrinet hardware
and the corresponding ParaStationII and GM drivers are sufficient for 
development purposes. Still we would not yet recommend to use them in 
production systems that demand uptimes of several months. Note that the
parallel version of FORM using the message passing is actually independent
of the type of network hardware and a particular implementation of MPI
or PVM. Therefore it can be used on any parallel system that provides
one of these message passing libraries.

\noindent
Since the parallel program is actually based on FORM version 3.0, 
which is in preparation and offers some new and powerful features, 
we will investigate whether and how these new features can be 
implemented in the parallel version. 
We are also interested in porting the parallel version of FORM to SMP 
(symmetric multi-processing) architectures, where directly IPC (inter-process 
communication) or threads can be used to minimize communication overhead.
The main aim of all these efforts is to get from the current stage of a 
stable prototype to an easy to use, powerful and reliable program that
is not an end in itself, but a useful tool in real life applications on a 
wide spectrum of (parallel) architectures.

\vspace{5mm}
\noindent
{\large\bf Acknowledgements}
\vspace{5mm}

\noindent
This paper was supported by the DFG-Graduiertenkolleg 
"Elementarteilchenphysik an Beschleunigern" and the 
DFG-Forschergruppe "Quantenfeldtheorie, Computer-Algebra 
und Monte-Carlo-Simulation" under contract number FOR 264/2-1.

\end{document}